\documentclass[aps,pre,reprint,superscriptaddress,nobibnotes]{revtex4-2}

\usepackage{dcolumn}

\usepackage{bm} 
\usepackage{graphicx} 
\usepackage{color}
\usepackage{subfigure}
\usepackage{hyperref}
\usepackage{latexsym}
\usepackage{amsthm}
\usepackage{amssymb}
\usepackage{verbatim}
\usepackage{wasysym}
\usepackage{threeparttable}
\usepackage{tabularx}

\DeclareGraphicsExtensions{.jpg, .pdf, .mps, .png, .eps, .ps, .EPS, .gif}
\DeclareRobustCommand{\crosscircle}{%
  \mathbin{\ooalign{%
    \hfil$\bigcirc$\hfil\cr
    \hfil\raisebox{-0.6ex}{\scalebox{2.00}{$\times$}}\hfil\cr
  }}%
}

\begin{document}
\def\be{\begin{equation}}
\def\ee{\end{equation}}

\def\bc{\begin{center}}
\def\ec{\end{center}}
\def\bea{\begin{eqnarray}}
\def\eea{\end{eqnarray}}

\newcommand{\avg}[1]{\langle{#1}\rangle}
\newcommand{\Avg}[1]{\left\langle{#1}\right\rangle}

\newcommand{\cch}[1]{\left[#1\right]}
\newcommand{\chv}[1]{\left \{ #1\right \} }
\newcommand{\prt}[1]{\left(#1\right)}
\newcommand{\aver}[1]{\left\langle #1 \right\rangle}
\newcommand{\abs}[1]{\left| #1 \right|}

\def\ie{\textit{i.e.}}
\def\etal{\textit{et al.}}
\def\m{\vec{m}}
\def\G{\mathcal{G}}
\def\fig{FIG.}
\def\tab{TABLE }

\newcommand{\gin}[1]{{\bf\color{magenta}#1}}
\newcommand{\bobred}[1]{{\color{red}#1}}
\newcommand{\bob}[1]{#1}
\newcommand{\bobz}[1]{{\bf\color{magenta}#1}}

\title{Extended-range percolation on lattices with cubic and non-compact neighborhoods in three dimensions}

\author{Zhe Zhang}
\affiliation{School of Physics Science and Technology, Yangzhou University, Yangzhou 225009, China}

\author{Yuxuan Fan}
\affiliation{School of Physics Science and Technology, Yangzhou University, Yangzhou 225009, China}

\author{Yang Yang}
\affiliation{School of Material Sciences and Physics, China University of Mining and Technology, Xuzhou 221116, China}

\author{Yusong Tu}
\affiliation{School of Physics Science and Technology, Yangzhou University, Yangzhou 225009, China}

\author{Zhipeng Xun}
\email{zpxun@cumt.edu.cn}
\affiliation{School of Material Sciences and Physics, China University of Mining and Technology, Xuzhou 221116, China}

\date{\today}

\begin{abstract}
Using Monte Carlo simulations based on an efficient single-cluster growth algorithm, we investigate two types of three-dimensional extended-range percolation models, namely cubic neighborhoods and complex (non-compact) neighborhoods. We obtain high-precision percolation thresholds $p_c$ for approximately 100 distinct structures and further explore the relationship between these thresholds and the topology of the lattices, which includes the total coordination number $z$, the number of sites $z_i$ and these sites' distance to the central site $r_i$ of the given neighborhood in the $i$th coordination zone. Both bond and site percolation are analyzed for cubic neighborhoods on the simple cubic lattice. For bond percolation, the asymptotic value of $zp_c$ tends to Bethe-lattice behavior ($zp_c \sim 1$), and finite-$z$ correction is found to be consistent with $zp_{c} - 1 \sim a_{1}z^{-x}$, where $x = (d-1)/d$ given by Xu et al.\ [Phys.\ Rev.\ E \textbf{103}, 022127 (2021)]. For site percolation, the value of $zp_c$ varies non-monotonically, exhibiting an initial increase followed by a decrease with the coordination number $z$. Linear fitting leads to $zp_c \sim 2.5987(1)$ for large $z$, predicting a value of continuum percolation threshold of aligned cubes $\eta_c = 0.324838(12)$ using $zp_c = 2^{d} \eta_c$. We then examine site percolation for non-compact (or complex) neighborhoods extending to the fifth coordination zone on simple cubic, face centered cubic, and body centered cubic lattices. Through a comparison of the three indices of the coordination number $z$, the weighted distance $\xi = \sum_i z_i r_i^2 /i$ , and the weighted coordination number $\zeta = \sum_i z_i r_i$, we find that $\zeta$ is effective in removing the degeneracy (a given $z$ corresponds to several different percolation thresholds $p_c$). The results obtained show power-law dependence $p_c \propto (\zeta / K)^{-\gamma}$ with $\gamma_{\mathrm{SC}}^{\zeta} = 0.703(12)$, $\gamma_{\mathrm{FCC}}^{\zeta} = 0.728(10)$, and $\gamma_{\mathrm{BCC}}^{\zeta} = 0.699(12)$ for simple cubic, face centered cubic, and body centered cubic lattices, respectively, and $p_c \propto \zeta^{-\gamma}$ with $\gamma^{\zeta} = 0.719(9)$ independent of the underlying lattice. The number $K$ indicates lattice connectivity, that is, $K = 6$, $12$, and $8$ for the simple cubic, face centered cubic, and body centered cubic lattices, respectively.
\end{abstract}


\maketitle
\section{Introduction}
Percolation \cite{BroadbentHammersley57, StaufferAharony94} provides a fundamental framework for describing connectivity-driven phase transitions in disordered systems and plays an important role in statistical physics. Beyond its conceptual importance for understanding critical phenomena and universality, percolation serves as a paradigmatic example for studying the interplay among geometry, randomness, and collective behavior. In addition, percolation has been applied to a variety of scientific problems, including fluid flow through porous media \cite{BolandtabaSkauge11,MourzenkoThovertAdler11}, forest fires \cite{Henley93,GuisoniLoscarAlbano11} and epidemics spreading \cite{MooreNewman00,Ziff21}. 

\begin{figure}[htbp] 
\centering
\includegraphics[width=3.1in]{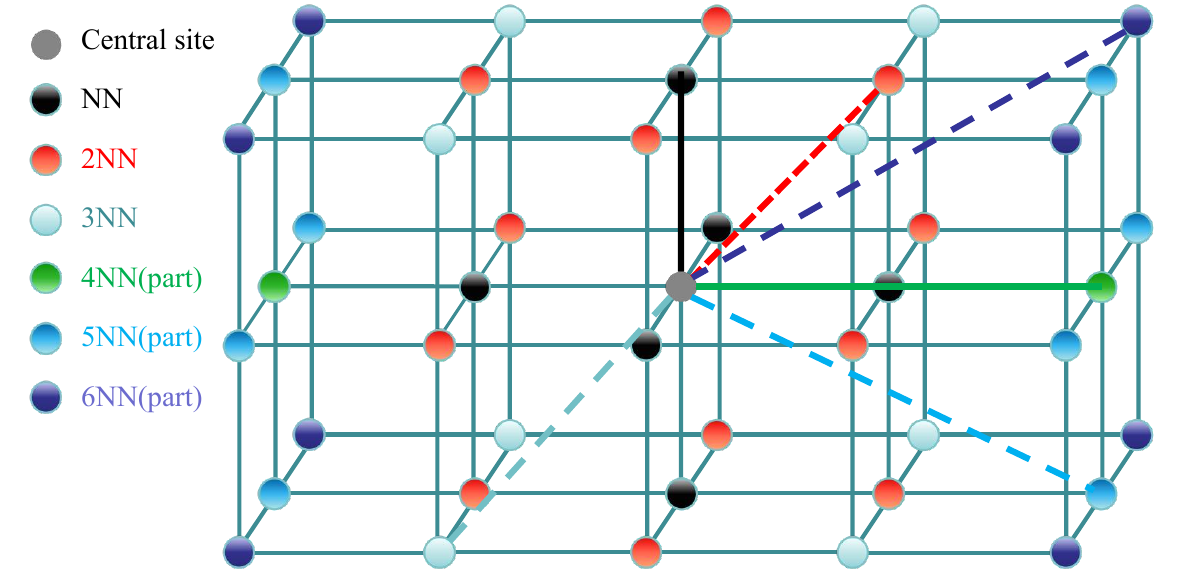}
\caption{Schematic illustration of extended-range neighborhoods on the simple cubic lattice. We use the notation \textsc{sc}-$a,b,...$ to indicate the lattice with the $a$th nearest neighbors, the $b$th nearest neighbors, etc. Suppose the farthest neighbor considered is 6NN, for compact neighborhood, all neighbors inside 6NN should be covered, i.e., \textsc{sc}-$1,2,3,4,5,6$; whereas a non-compact neighborhood includes only a subset of these shells, such as \textsc{sc}-$1,6$ and \textsc{sc}-$1,2,6$.}
\label{fig:diagram}
\end{figure}

Extended-range percolation, also referred to as equivalent neighbors percolation \cite{DombDalton66}, medium-range percolation \cite{FengDengBlote08}, or complex neighborhood percolation \cite{MalarzGalam05}, considers not only the nearest neighbors (NN) of a central site in a lattice model but also more distant neighbors, such as the second nearest neighbors (2NN), the third nearest neighbors (3NN), and so forth, as shown in Fig.\ \ref{fig:diagram}. Due to its close links to several important topics, such as problems of adsorption of extended shapes on a lattice \cite{KozaKondratSuszcaynski14,KozaPola16}, small-world networks \cite{Kleinberg00}, and the spread of epidemics \cite{Ziff21,SanderWarrenSokolov03}, extended-range percolation has long attracted considerable interest \cite{DaltonDombSykes64,GoukerFamily83,GawronCieplak91,KurzawskiMalarz12,Malarz15,KotwicaGronekMalarz19,Malarz20,Malarz21,Malarz22,Malarz23,Malarz24,MalarzKrawczyk25,CiepluchaUtnickiWoloszynMalarz25,OuyangDengBlote18,DengOuyangBlote19,XunZiff20b,XunHaoZiff21,XunHaoZiff22,ZhaoYanXunHaoZiff22,XunHaoZiff25,AmaralSoaresZiff26}. Moreover, this class of lattice models lies between discrete and continuum percolation, and further studies will help establish the relationship between these two problems \cite{Domb72,KozaKondratSuszcaynski14,KozaPola16}.

Two types of neighborhoods are usually considered in studies of extended-range percolation models, namely compact and non-compact neighborhoods, as illustrated in Fig.\ \ref{fig:diagram}. Previous studies of both neighborhood types have primarily focused on the relationship between the percolation threshold and the lattice topology. The percolation threshold $p_c$ separates the nonpercolating phase ($p < p_c$), in which the system behaves as an insulator, from the percolating phase ($p > p_c$), in which it behaves as a conductor. The lattice topology can be characterized by quantities such as the total coordination number $z$, the number of sites $z_i$ in the $i$th coordination zone, and their distance $r_i$ from the central site, etc \cite{XunHaoZiff22,Malarz24}.

With regard to the type of compact neighborhoods, for large $z$, one expects that Bethe-lattice behavior
\begin{equation}
    p_c = \frac{1}{z-1}
\label{eq:zpcbond}
\end{equation}
will hold for bond percolation, and it has been argued \cite{Penrose93,XunHaoZiff21} that  
\begin{equation}
    zp_c = 2^{d} \eta_{c}
\label{eq:zpcsite}
\end{equation}
for site percolation, where $\eta_c$ is the continuum percolation threshold for objects of the same shape as the neighborhood.

Many important results have been achieved in dimensions from $2$ to $5$, by means of numerical simulation. However, plots of $z p_c$ versus $1/z$ for site percolation in previous studies, such as the inset of Fig.\ 6 in Ref.\ \cite{XunHaoZiff25}, exhibit substantial scatter that cannot be attributed to statistical errors. This scatter was conjectured to arise from differences in neighborhood shape. In previous studies, the shape of the outer neighborhood shell often varied as the coordination number increases. This lack of geometric consistency may introduce uncontrolled artifacts. To address this issue, in the first part of this study, we restrict the neighborhoods to cubic shells so that the outer boundary remains cubic in all neighborhoods. This construction allows us to systematically increase $z$ while keeping the neighborhood geometry fixed.

For non-compact neighborhoods, the situation is more subtle. In two dimensions, distinct neighborhood compositions with the same $z$ have been observed to yield different percolation thresholds $p_c$, a phenomenon referred to as degeneracy \cite{Malarz24}. The connectivity structure depends not only on the total coordination number $z$ but also on the spatial distribution of the neighbors. To remove this $p_c(z)$ degeneracy in site percolation, Malarz introduced, in successive studies \cite{Malarz21,Malarz22,Malarz24}, the weighted square distance 
\begin{equation}
    \xi = \sum_i z_i r_i^2 /i
\label{eq:xiziri}
\end{equation}
and the weighted coordination number
\begin{equation}
    \zeta = \sum_i z_i r_i.
\label{eq:zetaziri}
\end{equation}
A power-law relation
\begin{equation}
    p_c \propto \xi^{-\gamma}
\label{eq:xi}
\end{equation}
was found for the triangular lattice \cite{Malarz21} with $\gamma ^ {\xi} _{\mathrm{TR}} \approx 0.710(19)$, whereas 
\begin{equation}
    p_c \propto \zeta^{-\gamma}
\label{eq:zeta}
\end{equation}
was obtained for the honeycomb lattice \cite{Malarz22} with $\gamma ^ {\zeta} _{\mathrm{HC}} \approx 0.4981(90)$. In a recent study, Malarz \cite{Malarz24} further reported the power-law dependence $p_c \propto (\zeta /K)^{-\gamma}$ with $\gamma^{\zeta} = 0.526(11)$, $0.5439(63)$, and $0.5932(47)$ for honeycomb, square, and triangular lattices, respectively, together with the lattice-independent relation $p_c \propto \zeta ^ {-\gamma}$ with $\gamma^{\zeta} = 0.5546(67)$. The number $K$ indicates lattice connectivity, that is, $K = 3$, $4$, and $6$ for the honeycomb, square, and triangular lattices, respectively.

The index $\xi$ was found to be redundant relative to $\zeta$ for the site percolation. However, $\xi$ was found to be quite useful for bond percolation, for which the relationship in Eq.\ (\ref{eq:xi}) is perfectly satisfied with the exponent $\gamma^{\xi} \approx 1$ \cite{XunHao22}. Furthermore, the research results obtained thus far are confined to two dimensions, leaving the open question of how the behavior extends to higher dimensions. To further compare the two indices $\xi$ versus $\zeta$ and, meanwhile, obtain universal power-law behavior across different dimensions, we extend the investigation to three dimensions in the second part of this study. We conducted extensive numerical simulations of complex nearest neighbor structures of simple cubic (SC), face centered cubic (FCC), and body centered cubic (BCC) lattices, with neighborhoods extending to the fifth-nearest-neighbors shell.

The remainder of the paper is organized as follows. In Sec.\ \ref{bk}, we briefly introduce the simulation method and scaling theory. Results for cubic and non-compact neighborhoods are presented in Secs.\ \ref{cubic} and \ref{non}, respectively, followed by the conclusions in Sec.\ \ref{conclusions}.

\section{Background Knowledge}
\label{bk}
\subsection{Simulation Method}
We consider bond and site percolation on a SC lattice with extended-range neighborhoods, as illustrated in Fig.\ \ref{fig:diagram}. The lattice constant is set to unity. An extended neighborhood with a cubic outer boundary, hereafter referred to as a cubic neighborhood, is denoted by \textsc{cub}-$(c)$, where $c$ represents the linear size of its outer boundary. Throughout this paper, the notation \textsc{cub}-$(c)$ is used exclusively for cubic neighborhoods on the SC lattice. For a given \textsc{cub}-$(c)$, the coordination number $z$ is given by
\begin{equation}
    z = (c + 1) ^ {3} - 1,
\label{eq:z}
\end{equation}
where $c$ should be even to ensure that the central site is located at the center of the cube. 

For non-compact neighborhoods, we use the notation \textsc{sc}-$a,b,...$ to indicate a SC lattice whose neighborhood includes the $a$th- and $b$th- nearest neighbor shell, etc. Similar notations are used to designate FCC and BCC lattices. Then the coordination number $z$ is given by
\begin{equation}
    z(\textsc{sc}-a,b,\ldots) = z(\textsc{sc}-a)+z(\textsc{sc}-b)+\cdots.
\label{eq:znon}
\end{equation}

To determine the percolation threshold $p_c$, we employ extensive Monte Carlo simulations based on a single-cluster growth algorithm. Simulations are performed on $L \times L \times L$ lattices with $L = 2^{10} (1024)$, corresponding to a system volume $V = L^3 = 2^{30}$. Periodic boundary conditions are applied in all three spatial directions. 

For each lattice, we perform at least $10^7$ independent cluster-growth realizations, with the exact number specified in the corresponding table. In each realization, a seeded site is chosen randomly, and a cluster grows according to the occupation probability $p$. The cluster size $s$ is defined as the total number of connected occupied sites (or bonds, depending on the model). Cluster sizes are collected into logarithmic bins of the form $[2^n,2^{n+1})$, with $n=0,1,2,\ldots$. To ensure that the clusters do not wrap around the lattice, we impose an upper cutoff on the cluster size. The maximum cluster size is restricted to $s_{\mathrm{max}} = 2^{16}$ for $z < 3000$, and for $z \ge 3000$, we set the maximum cluster size to $s_{\mathrm{max}} = 2^{20}$, which is still much smaller than the system volume. Clusters still growing when they reach the upper size cutoff are counted in the last bin. A more detailed description of our algorithm is given in Refs.\ \cite{XunZiff20,XunZiff20b,XunHaoZiff21}.

\subsection{Scaling Theory}
The statistical properties of clusters are usually characterized by the cluster number density $n_s(p)$, which is defined as the number of clusters per lattice site containing exactly $s$ occupied sites or bonds at occupation probability $p$. At the percolation threshold $p_c$, $n_s(p)$ is expected to follow the scaling form
\begin{equation}
n_s \sim A_0 s^{-\tau}\left(1 + B_0 s^{-\Omega} + \cdots \right),
\label{eq01}
\end{equation}
where $\tau$ is the Fisher exponent and $\Omega$ is the leading correction-to-scaling exponent. 
The exponents $\tau$ and $\Omega$ are expected to be universal, meaning that they take the same values for all lattices and percolation types in a given dimension. The amplitudes $A_0$ and $B_0$ are nonuniversal constants that depend on lattice geometry.

The probability that a randomly chosen site (or bond) belongs to a cluster of size greater than or equal to $s$ is given by
\begin{equation}
    P_{\ge s} = \sum_{s'=s}^{\infty} s' n_{s'}.
\label{eq02}
\end{equation}
At $p = p_c$, substituting Eq.\ (\ref{eq01}) into Eq.\ (\ref{eq02}) yields
\begin{equation}
    P_{\ge s} \sim A_1 s^{2-\tau}\left(1 + B_1 s^{-\Omega} + \cdots \right),
\label{eq03}
\end{equation}
where $A_1 = A_0/(\tau-2)$ and $B_1 = (\tau-2)B_0/(\tau+\Omega-2)$. Thus
\begin{equation}
    s^{\tau-2}P_{\ge s} \sim A_1 \left(1 + B_1 s^{-\Omega} + \cdots \right).
\label{eq:omega}
\end{equation}

When the probability $p$ differs from $p_{c}$, $P_{\ge s}$ can be written, using the universal exponent $\sigma$, as
\begin{equation}
    P_{\ge s} \sim A_2 s^{2-\tau} f\!\left[ B_2 \left(p-p_c\right) s^{\sigma} \right],
\label{eq04}
\end{equation}
where $f(x)$ is a scaling function. In the scaling limit $s \rightarrow \infty$ and $p \rightarrow p_{c}$, the scaling function $f(x)$ can be expanded as a Taylor series
\begin{equation}
    f(B_2(p-p_{c})s^{\sigma}) \sim 1+C_2(p-p_{c})s^{\sigma}+ \cdot\cdot\cdot,
\label{scaling}
\end{equation}
where $C_2 = B_2 f'(0)$. Assuming $f(0)=1$ gives $A_2$ = $A_1$. We then obtain
\begin{equation}
    s^{\tau-2} P_{\ge s} \sim A_2 + D_2 \left(p-p_c\right) s^{\sigma},
\label{eq06}
\end{equation}
where $D_2=A_2 C_2$.

As a consequence, Eq.\ (\ref{eq06}) implies that the quantity $s^{\tau-2} P_{\ge s}$ approaches a constant value for large $s$ when $p=p_c$. In contrast, for $p \neq p_c$, this quantity exhibits systematic deviations from a constant behavior. This property provides a practical criterion for locating the percolation threshold. In our simulations, we use $\sigma \approx 0.4522$ \cite{BallesterosFernandezMartin-MayorSudupeParisiRuiz-Lorenzo97} and $\tau \approx 2.18905$, as in our previous work \cite{XunZiff20b}.

\section{Results for cubic neighborhoods}
\label{cubic}
\subsection{Bond percolation}
\begin{figure*}[htbp]
\centering
\includegraphics[width=6.9in]{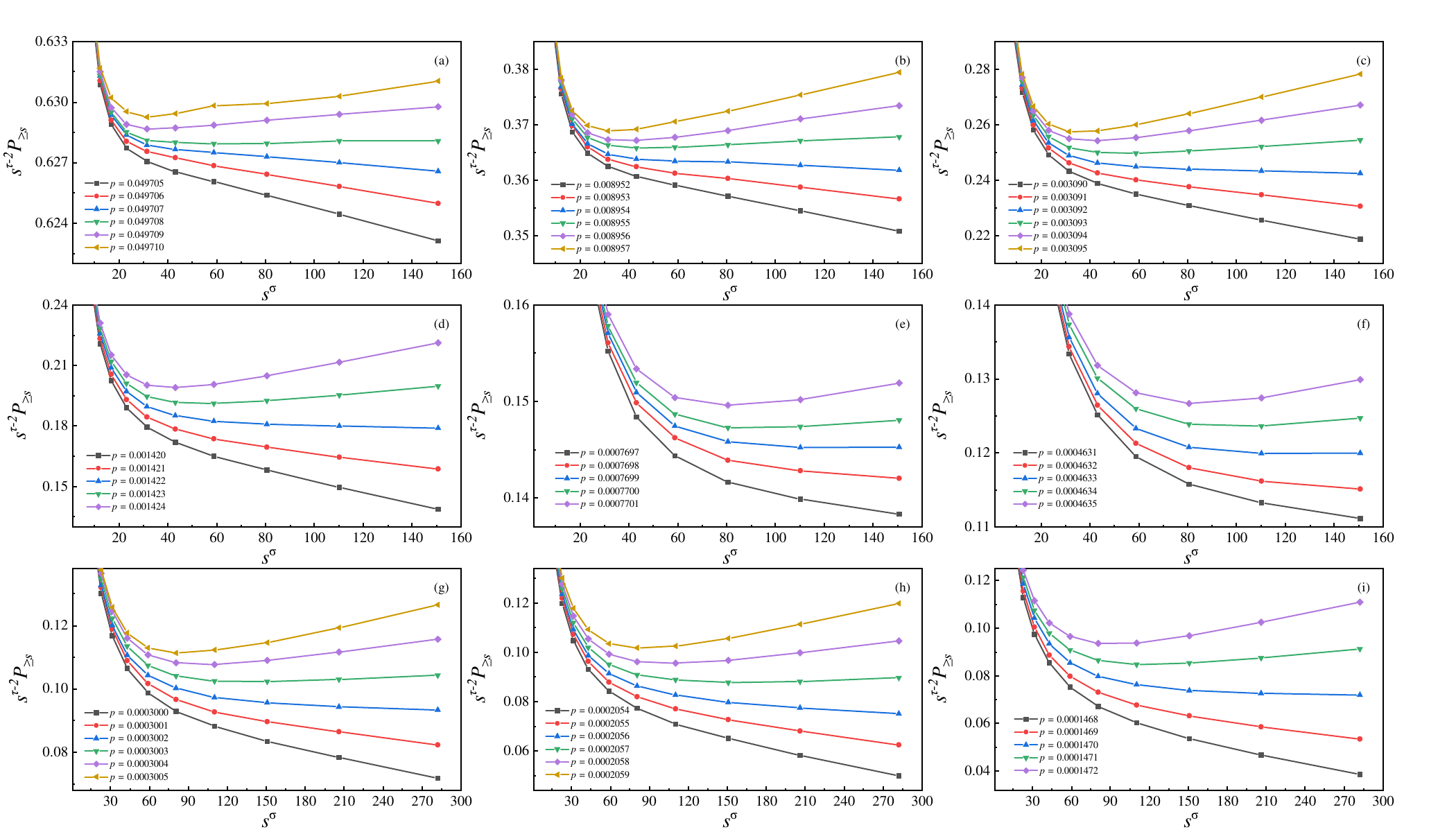}
\caption{Plots of $s^{\tau-2}P_{\geq{s}}$ versus $s^{\sigma}$ with $\tau=2.18905$ and $\sigma=0.4522$ for bond percolation on the \textsc{cub}-$(2)$ (a), \textsc{cub}-$(4)$ (b), \textsc{cub}-$(6)$ (c), \textsc{cub}-$(8)$ (d), \textsc{cub}-$(10)$ (e), \textsc{cub}-$(12)$ (f), \textsc{cub}-$(14)$ (g), \textsc{cub}-$(16)$ (h), and \textsc{cub}-$(18)$ (i) neighborhoods for different values of $p$. For each neighborhood, as $p \rightarrow p_c$, the value of $s^{\tau-2}P_{\geq{s}}$ becomes horizontal for large $s$.} 
\label{bond_sigma_33} 
\end{figure*}

Figure \ref{bond_sigma_33} presents the plots of $s^{\tau-2}P_{\ge s}$ versus $s^{\sigma}$ for bond percolation on a SC lattice with cubic neighborhoods $\textsc{cub}$-$(c)$, where $c = 2, 4, \ldots, 18$. At small cluster sizes, the curves exhibit appreciable curvature arising from non-asymptotic corrections and finite-size effects. As $s$ increases, linear behavior gradually begins to emerge.

\begin{figure*}[htbp]
\centering
\includegraphics[width=6.9in]{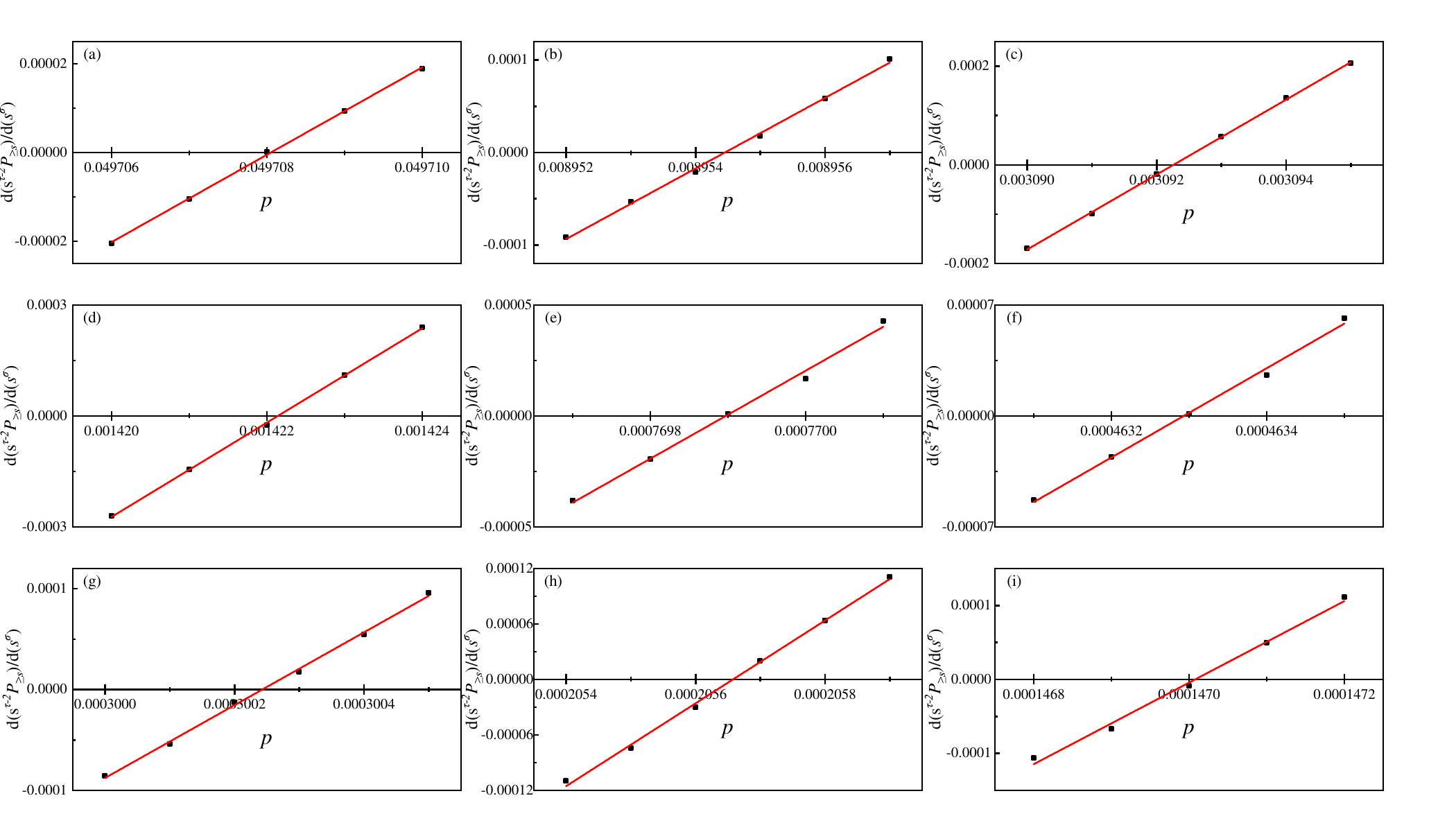}
\caption{Plots of $\mathrm{d}(s^{\tau-2}P_{\geq{s}})/\mathrm{d}(s^{\sigma})$ versus $p$ for bond percolation on the \textsc{cub}-$(2)$ (a), \textsc{cub}-$(4)$ (b), \textsc{cub}-$(6)$ (c), \textsc{cub}-$(8)$ (d), \textsc{cub}-$(10)$ (e), \textsc{cub}-$(12)$ (f), \textsc{cub}-$(14)$ (g), \textsc{cub}-$(16)$ (h), and \textsc{cub}-$(18)$ (i) neighborhoods. The central value of $p_c$ can be estimated from the $p$-axis intercept.}
\label{bond_sigma_diff_33} 
\end{figure*}

For $p < p_c$, the large-$s$ portion of the curve has a negative slope, while for $p>p_c$, the slope becomes positive. The value of $p$ for which the large-$s$ portion is closest to horizontal therefore provides an initial estimate of $p_c$. This visual estimate can be quantified by plotting $\mathrm{d}(s^{\tau-2}P_{\geq s}) / \mathrm{d}(s^{\sigma})$ versus $p$. Then the central value of percolation threshold $p_c$ can be determined from the $p$-axis intercept. Figure \ref{bond_sigma_diff_33} shows the derivative $\mathrm{d}(s^{\tau-2}P_{\geq s}) / \mathrm{d}(s^{\sigma})$ as a function of the occupation probability $p$ for bond percolation on the \textsc{cub}-$(c)$ objects with $c = 2, 4, \dots, 18$. The final percolation thresholds obtained are summarized in the fourth column of Table \ref{tab01}, and the numbers in parentheses represent the estimated uncertainties in the last digit. 

\begin{table}[htb]
	\caption{Bond percolation thresholds for each \textsc{cub}-$(c)$ neighborhood.}
	\begin{tabular}{ccccc}
	\hline\hline
		neighborhood         & runs            & $z$      & $p_c$                    & $zp_c$ \\
		\hline
		\textsc{cub}-$(2)$   & $8\times 10^8$  & $26$     & $0.0497080(10)$          & $1.2924$ \\
		\textsc{cub}-$(4)$   & $1\times 10^8$  & $124$    & $0.0089544(6)$           & $1.1103$ \\
		\textsc{cub}-$(6)$   & $1\times 10^8$  & $342$    & $0.0030923(7)$           & $1.0576$ \\
		\textsc{cub}-$(8)$   & $1\times 10^8$  & $728$    & $0.0014223(7)$           & $1.0354$ \\
		\textsc{cub}-$(10)$  & $1\times 10^8$  & $1330$   & $0.00076989(9)$          & $1.0240$ \\
		\textsc{cub}-$(12)$  & $7\times 10^7$  & $2196$   & $0.00046332(8)$          & $1.0175$ \\
		\textsc{cub}-$(14)$  & $3\times 10^7$  & $3374$   & $0.00030032(8)$          & $1.0133$ \\
		\textsc{cub}-$(16)$  & $3\times 10^7$  & $4912$   & $0.00020572(8)$          & $1.0105$ \\
		\textsc{cub}-$(18)$  & $2\times 10^7$  & $6858$   & $0.00014707(7)$          & $1.0086$ \\
	\hline\hline
	\end{tabular}
        \label{tab01}
\end{table}	

Besides the percolation thresholds, Table \ref{tab01} also lists the corresponding coordination number $z$ and the product $z p_c$ for each neighborhood. A clear monotonic decrease of the percolation threshold $p_c$ with increasing coordination number $z$ is observed. As the neighborhood range expands from \textsc{cub}-$(2)$ to \textsc{cub}-$(18)$, the coordination number $z$ increases from 26 to 6858, while the threshold $p_c$ decreases by more than two orders of magnitude.

\begin{figure}[htbp]
\centering
\includegraphics[width=3.1in]{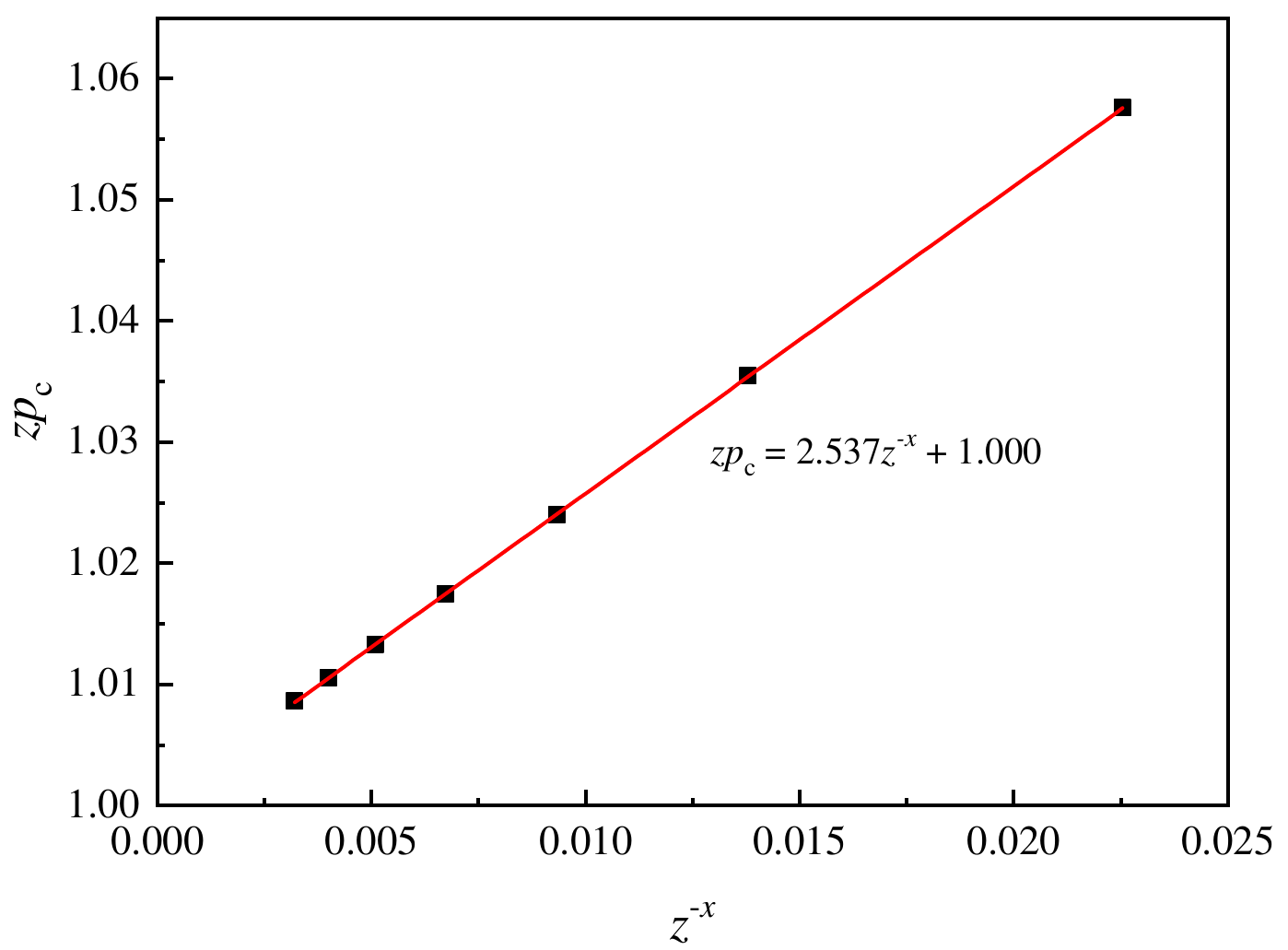}
\caption{Plot of $zp_c$ versus $z^{-x}$ for bond-percolation data listed in Table \ref{tab01} with $z \geq 342$. The choice $x = 0.65$ gives a good linear fit, $z p_c \sim 1.000 + 2.537 z^{-0.65}$, with an intercept close to $1$.}
\label{bondzpc}
\end{figure}

More importantly, the quantity $z p_c$ exhibits a systematic convergence toward unity as $z$ increases. Starting from $z p_c \approx 1.2924$ for \textsc{cub}-$(2)$, the value decreases monotonically and approaches $z p_c \approx 1.0086$ for \textsc{cub}-$(18)$. With regard to finite $z$, theoretical analysis for bond thresholds has been given by Xu et al \cite{XuWangHuDeng21} as
\begin{equation}
    zp_{c} - 1 \sim a_{1}z^{-x},
\label{eq:zpcbond2}
\end{equation}
for $d=2$ and 3, where $x = (d-1)/d$.
For $z \ge 342$, the fit of $z p_c$ versus $z^{-x}$ is shown in Fig.\ \ref{bondzpc}. A linear fit yields
\begin{equation}
  z p_c = 1.000 + 2.537z^{-0.65}. 
  \label{eq_bond_zpc}
\end{equation}
Thus, choosing $x = 0.65$ gives a good linear fit and an intercept close to the asymptotic value of 1, corresponding to Bethe-lattice behavior. The value of $x$ obtained here is consistent with the theoretical prediction $x = 2/3$ of Xu et al \cite{XuWangHuDeng21}.

\subsection{Site percolation}
\label{sec:site}
Now we turn to the results of site percolation. Using the same scaling framework employed for bond percolation, the plots of $s^{\tau-2} P_{\ge s}$ versus $s^{\sigma}$ for site percolation on the \textsc{cub}-$(2)$ through \textsc{cub}-$(22)$ neighborhoods at different values of the occupation probabilities $p$ are shown in Fig.\ S1 of the Supplemental Material \cite{ZhangFanYangXun26supplementary}. As in bond percolation, the curves tend to become horizontal at large $s$ as $p$ approaches the threshold $p_c$.

Figure S2 in the Supplemental Material \cite{ZhangFanYangXun26supplementary} presents the plots of $\mathrm{d}(s^{\tau-2}P_{\ge s})/\mathrm{d}(s^{\sigma})$ as a function of the occupation probability $p$ for site percolation on the \textsc{cub}-$(2)$ through \textsc{cub}-$(22)$ objects. For each neighborhood size, this quantity exhibits a linear dependence on $p$ in the vicinity of the critical point. The values of $p_c$ can therefore be identified reliably from the intercepts, and the estimated thresholds are summarized in Table \ref{tab02}.

\begin{table}[htb]
	\caption{Site percolation thresholds for each \textsc{cub}-$(c)$ neighborhood.}
	\begin{threeparttable}
        \begin{tabular}{ccccc}
	\hline\hline
		neighborhood         & runs           & $z$     & $p_c$                  & $zp_c$ \\
		\hline
		\textsc{cub}-$(2)$   & $1\times 10^8$ & $26$    & $0.097644(6)$\tnote{a} & $2.5387$ \\
		\textsc{cub}-$(4)$   & $1\times 10^8$ & $124$   & $0.021037(7)$          & $2.6086$ \\
		\textsc{cub}-$(6)$   & $1\times 10^8$ & $342$   & $0.0076327(7)$         & $2.6104$ \\
		\textsc{cub}-$(8)$   & $1\times 10^8$ & $728$   & $0.0035822(8)$         & $2.6078$ \\
		\textsc{cub}-$(10)$  & $1\times 10^8$ & $1330$  & $0.0019592(8)$         & $2.6057$ \\
		\textsc{cub}-$(12)$  & $7\times 10^7$ & $2196$  & $0.0011861(9)$         & $2.6047$ \\
		\textsc{cub}-$(14)$  & $3\times 10^7$ & $3374$  & $0.00077142(8)$        & $2.6028$ \\
		\textsc{cub}-$(16)$  & $3\times 10^7$ & $4912$  & $0.00052971(9)$        & $2.6019$ \\
		\textsc{cub}-$(18)$  & $2\times 10^7$ & $6858$  & $0.00037934(6)$        & $2.6015$ \\
        \textsc{cub}-$(20)$  & $1\times 10^7$ & $9260$  & $0.00028090(10)$       & $2.6011$ \\
        \textsc{cub}-$(22)$  & $1\times 10^7$ & $12166$ & $0.00021378(8)$        & $2.6008$ \\
	\hline\hline
	\end{tabular}
    \begin{tablenotes}
    \footnotesize
        \item[a] $0.0976445(10)$ and $0.0976444(6)$\cite{XunHaoZiff21}, and $0.0976(1)$\cite{KurzawskiMalarz12}.
    \end{tablenotes}
	\label{tab02}
    \end{threeparttable}
\end{table}	

\begin{figure}[htbp]
\centering
\includegraphics[width=3.1in]{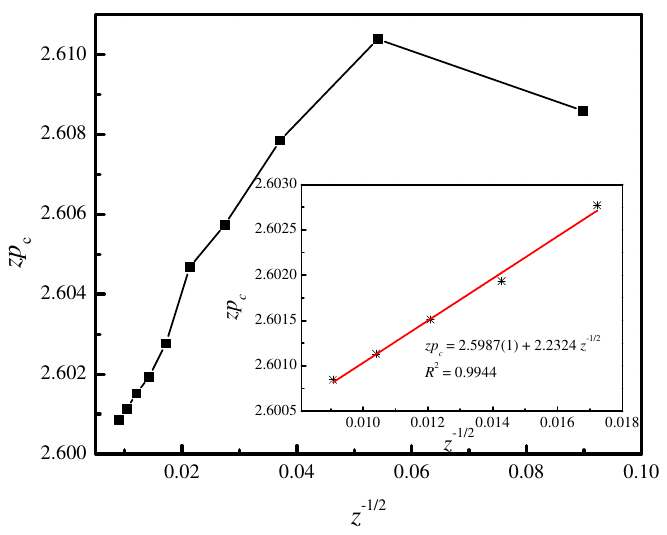}
\caption{Plot of $zp_c$ versus $z^{-1/2}$ for site-percolation data listed in Table \ref{tab02} with $z \ge 124$. The value of $zp_c$ exhibits an initial increase followed by a decrease with $z$. Furthermore, for small $z$, the variation is nonlinear, and a good linear relationship is observed only when $z$ is large. The inset shows a linear fit for $z \ge 3374$, yielding an intercept of $2.5987(1)$ and a slope of $2.2324$.}
\label{sitezpc}
\end{figure}

As shown in Table \ref{tab02}, the value of $zp_c$ varies non‑monotonically with the coordination number $z$, increasing initially and then decreasing, a behavior that differs from that reported in most previous studies. Motivated by the $\ell^{-3/2}$ convergence toward the continuum limit reported for aligned hypercubes \cite{KozaPola16}, and noting that $z \propto \ell^{3}$ for the present three-dimensional cubic neighborhoods, we plot $zp_c$ as a function of $z^{-1/2}$ for $z \ge 124$, as shown in Fig.\ \ref{sitezpc}. 

In contrast to the bond percolation case, where $z p_c$ exhibits a nearly linear approach toward the Bethe-lattice limit, the data of site percolation display a clear nonlinear dependence on $z^{-1/2}$ at the beginning. A satisfactory linear correlation is observed only when $z$ reaches a large value. For $z \ge 3374$, a linear fit to the data in the inset of Fig.\ \ref{sitezpc} yields an intercept of $2.5987(1)$. Then using the relation $zp_c = 2^{d} \eta_c$ from Eq. (\ref{eq:zpcsite}), we predict a value of continuum percolation threshold for aligned cubes to be $\eta_c = 0.324838(12)$. This value can be compared with existing results of $0.3247(3)$ \cite{TorquatoJiao12}, $0.3248(3)$ \cite{BakerPaulSreenivasanStanley02}, $0.32476(4)$ \cite{HyytiaVirtamoLassilaOtt12}, and $0.324766(1)$ \cite{KozaPola16}.

Although the variation is nonlinear over a wide range, the curve is smooth and does not exhibit the scatter observed in the inset of Fig.\ 6 in Ref.\ \cite{XunHaoZiff25}. We conjectured that the data scatter reported in Ref.\ \cite{XunHaoZiff25} arose from differences in the neighborhood shape. Here, we keep the neighborhood shape fixed and find that the large data scatter disappears. This confirms our previous speculation about the effects of differing shapes.

\section{Results for non-compact neighborhoods}
\label{non}

For non-compact (or complex) neighborhoods, we focus on the site percolation. Extensive numerical simulations were carried out for complex neighborhoods that extend to the fifth nearest neighbors on SC, FCC, and BCC lattices. The same procedure as that described in Section \ref{cubic} was applied. The plots of $s^{\tau-2} P_{\ge s}$ versus $s^{\sigma}$ and $\mathrm{d}(s^{\tau-2}P_{\ge s})/\mathrm{d}(s^{\sigma})$ versus $p$ for site percolation with complex neighborhoods on the SC, FCC, and BCC lattices are shown in Figs.\ S3-S6, Figs.\ S7-S10, and Figs.\ S11-S14 of the Supplemental Material \cite{ZhangFanYangXun26supplementary}, respectively. The corresponding estimated percolation thresholds are summarized in Tables \ref{tab03} (SC), \ref{tab04} (FCC), and \ref{tab05} (BCC), respectively.

Among the numerous neighborhood structures, some are equivalent to each other, such as \textsc{sc}-$2$ and \textsc{fcc}-$1$, \textsc{sc}-$1,2$, \textsc{fcc}-$2,4$ and \textsc{bcc}-$2,3$, etc. More detailed equivalence relations are listed in the footnotes of Table \ref{tab03}. Besides, the comparisons of our values with existing percolation thresholds are also described in the footnotes of each table. It is observable that some of our simulation results show a certain degree of discrepancy from previously reported results, such as those denoted by footnotes $l$, $m$, and $n$ in Table \ref{tab03}. We consider our simulation results to be more reliable. A case in point is that the percolation threshold reported in Ref.\ \cite{Malarz15} for the \textsc{sc}‑$2,3,4$ structure $0.11330(12)$ is higher than that reported in Ref.\ \cite{KurzawskiMalarz12} for the \textsc{sc}‑$2,3$ structure $0.1036(1)$. This ordering is inconsistent with the neighborhood-inclusion relation because \textsc{sc}‑$2,3,4$ contains all the nearest‑neighbor sites of \textsc{sc}‑$2,3$ and should therefore have a lower threshold.

Strong degeneracy, in which a given $z$ corresponds to several different percolation thresholds $p_c$, is evident in Tables \ref{tab03}, \ref{tab04} and \ref{tab05}. For example, in Table \ref{tab03}, $z(\textsc{sc}-1,2) = z(\textsc{sc}-2,4) =18$, whereas $p_c = 0.13730(10)$ for the \textsc{sc}-$1,2$ neighborhood and $p_c = 0.13614(6)$ for the \textsc{sc}-$2,4$ neighborhood. In this context, $z$ is not a suitable index for differentiating between various structures. The weighted distance $\xi$ defined in Eq.\ (\ref{eq:xiziri}) and the weighted coordination number $\zeta$ defined in Eq.\ (\ref{eq:zetaziri}) were also calculated. Their values are listed in columns 3 and 4 in Tables \ref{tab03}, \ref{tab04}, and \ref{tab05}. For the SC and FCC lattices, the normalized distances of the first five coordination zones satisfy $r_i^2=i$. Equation (\ref{eq:xiziri}) therefore gives $\xi=\sum_i z_i r_i^2/i=\sum_i z_i=z$; thus, $\xi$ cannot remove the degeneracy for these two lattices, although it has proved useful in two dimensions and for the BCC lattice in three dimensions [see Table \ref{tab05}]. In contrast to the first two indices, the weighted coordination number $\zeta$ effectively removes the degeneracy. 

\begin{table}[htbp]
\centering
\caption{Site percolation thresholds for objects of various complex neighborhoods on simple cubic lattice. The object name encodes the involved coordination zones ($i\le5$), to which sites in the neighborhood belong. The coordination number $z$, the weighted distance $\xi$, and the weighted coordination number $\zeta$ are also presented. Note that $\xi = z$.}
    \begin{threeparttable}
	\begin{tabular}{ccccc}
	\hline\hline
		neighborhood                  & $z$      &$\xi$      & $\zeta$    & $p_c$(site) \\
		\hline
		\textsc{sc}-$1$\tnote{aa}     & $6$      & $6$       & $6$        & $0.31161(9)$\tnote{a} \\
		\textsc{sc}-$2$\tnote{bb}     & $12$     & $12$      & $16.9706$  & $0.19924(6)$\tnote{b} \\
        \textsc{sc}-$1,2$\tnote{cc}   & $18$     & $18$      & $22.9706$  & $0.13730(10)$\tnote{c} \\
        \textsc{sc}-$3$\tnote{dd}     & $8$      & $8$       & $13.8564$  & $0.24596(6)$\tnote{d} \\
        \textsc{sc}-$1,3$\tnote{ee}   & $14$     & $14$      & $19.8564$  & $0.13615(5)$\tnote{e} \\
        \textsc{sc}-$2,3$\tnote{ff}   & $20$     & $20$      & $30.8270$  & $0.10376(6)$\tnote{f} \\
        \textsc{sc}-$1,2,3$\tnote{gg} & $26$     & $26$      & $36.8270$  & $0.097644(6)$\tnote{g} \\
        \textsc{sc}-$4$               & $6$      & $6$       & $12$       & $0.31161(9)$ \\
        \textsc{sc}-$1,4$             & $12$     & $12$      & $18$       & $0.15038(8)$\tnote{h} \\
        \textsc{sc}-$2,4$\tnote{hh}   & $18$     & $18$      & $28.9706$  & $0.13614(6)$\tnote{i} \\
        \textsc{sc}-$3,4$\tnote{ii}   & $14$     & $14$      & $25.8564$  & $0.17594(6)$\tnote{j} \\
        \textsc{sc}-$1,2,4$           & $24$     & $24$      & $34.9706$  & $0.099663(7)$\tnote{k} \\
        \textsc{sc}-$1,3,4$           & $20$     & $20$      & $31.8564$  & $0.10388(8)$\tnote{l} \\
        \textsc{sc}-$2,3,4$           & $26$     & $26$      & $42.8270$  & $0.085648(8)$\tnote{m} \\
        \textsc{sc}-$1,2,3,4$         & $32$     & $32$      & $48.8270$  & $0.080117(7)$\tnote{n} \\
        \textsc{sc}-$5$               & $24$     & $24$      & $53.6656$  & $0.067975(5)$ \\
        \textsc{sc}-$1,5$             & $30$     & $30$      & $59.6656$  & $0.061377(7)$ \\
        \textsc{sc}-$2,5$             & $36$     & $36$      & $70.6362$  & $0.057204(6)$ \\
        \textsc{sc}-$3,5$             & $32$     & $32$      & $67.5220$  & $0.056993(7)$ \\
        \textsc{sc}-$4,5$             & $30$     & $30$      & $65.6656$  & $0.059547(7)$ \\
        \textsc{sc}-$1,2,5$           & $42$     & $42$      & $76.6362$  & $0.054334(6)$ \\
        \textsc{sc}-$1,3,5$           & $38$     & $38$      & $73.5220$  & $0.053423(7)$ \\
        \textsc{sc}-$1,4,5$           & $36$     & $36$      & $71.6656$  & $0.056210(10)$ \\
        \textsc{sc}-$2,3,5$           & $44$     & $44$      & $84.4926$  & $0.051515(5)$ \\
        \textsc{sc}-$2,4,5$           & $42$     & $42$      & $82.6362$  & $0.051596(6)$ \\
        \textsc{sc}-$3,4,5$           & $38$     & $38$      & $79.5220$  & $0.051583(7)$ \\
        \textsc{sc}-$1,2,3,5$         & $50$     & $50$      & $90.4926$  & $0.049696(6)$ \\
        \textsc{sc}-$1,2,4,5$         & $48$     & $48$      & $88.6362$  & $0.050017(7)$ \\
        \textsc{sc}-$1,3,4,5$         & $44$     & $44$      & $85.5220$  & $0.049482(8)$ \\
        \textsc{sc}-$2,3,4,5$         & $50$     & $50$      & $96.4926$  & $0.047257(7)$ \\
        \textsc{sc}-$1,2,3,4,5$       & $56$     & $56$      & $102.4926$ & $0.046182(8)$\tnote{o} \\
	\hline\hline
	\end{tabular}
    \begin{tablenotes}
    \footnotesize
    \item[aa] Equivalent to \textsc{sc}-$4$, \textsc{fcc}-$2$, \textsc{bcc}-$2$.
    \item[bb] Equivalent to \textsc{fcc}-$1$, \textsc{fcc}-$4$, \textsc{bcc}-$3$.
    \item[cc] Equivalent to \textsc{fcc}-$2,4$, \textsc{bcc}-$2,3$.
    \item[dd] Equivalent to \textsc{bcc}-$1$, \textsc{bcc}-$5$.
    \item[ee] Equivalent to \textsc{bcc}-$2,5$.
    \item[ff] Equivalent to \textsc{bcc}-$3,5$.
    \item[gg] Equivalent to \textsc{cub}-$(2)$, \textsc{bcc}-$2,3,5$.
    \item[hh] Equivalent to \textsc{fcc}-$1,2$.
    \item[ii] Equivalent to \textsc{bcc}-$1,2$. 
    \item[a] $0.31160768(15)$ \cite{XuWangLvDeng14}, $0.3116060(48)$ \cite{KozaPola16}, $0.3116077(2)$ \cite{WangZhouZhangGaroniDeng13}, $0.3116081(13)$ \cite{BallesterosFernandezMartin-MayorParisiRuiz-Lorenzo99}.
    \item[b] $0.1991(1)$ \cite{KurzawskiMalarz12}.
    \item[c] $0.1373045(5)$ \cite{XunHaoZiff21}, $0.1372(1)$ \cite{KurzawskiMalarz12}.
    \item[d] $0.2455(1)$ \cite{KurzawskiMalarz12}.
    \item[e] $0.1420(1)$ \cite{KurzawskiMalarz12}.
    \item[f] $0.1036(1)$ \cite{KurzawskiMalarz12}.
    \item[g] $0.0976445(10)$, $0.0976444(6)$ \cite{XunHaoZiff21}, $0.0976(1)$ \cite{KurzawskiMalarz12}.
    \item[h] $0.1503793(7)$ \cite{XunHaoZiff21}, $0.15040(12)$ \cite{Malarz15}.
    \item[i] $0.15950(12)$ \cite{Malarz15}.
    \item[j] $0.20490(12)$ \cite{Malarz15}.
    \item[k] $0.11440(12)$ \cite{Malarz15}.
    \item[l] $0.11920(12)$ \cite{Malarz15}.
    \item[m] $0.11330(12)$ \cite{Malarz15}.
    \item[n] $0.10000(12)$ \cite{Malarz15}, $0.0801171(9)$ \cite{XunHaoZiff21}.
    \item[o] $0.0461815(5)$ \cite{XunHaoZiff21}.
    \end{tablenotes}
	\label{tab03}
    \end{threeparttable}
\end{table}	

\begin{table}[htbp]
\centering
\caption{Site percolation thresholds for objects of various complex neighborhoods on face centered cubic lattice. The object name encodes the involved coordination zones ($i\le5$), to which sites in the neighborhood belong. The coordination number $z$, the weighted distance $\xi$, and the weighted coordination number $\zeta$ are also presented. Here $r$ was normalized to ensure that $r_1 = 1$. Note that $\xi = z$.}
    \begin{threeparttable}
	\begin{tabular}{ccccc}
	\hline\hline
		neighborhood                & $z$     & $\xi$     & $\zeta$    & $p_c$(site) \\
		\hline
		\textsc{fcc}-$1$            & $12$    & $12$      & $12$  & $0.19924(6)$\tnote{a} \\
		\textsc{fcc}-$2$            & $6$     & $6$      & $8.4853$       & $0.31161(9)$\tnote{b} \\
        \textsc{fcc}-$1,2$          & $18$    & $18$      & $20.4853$  & $0.13614(6)$\tnote{c} \\        
        \textsc{fcc}-$3$            & $24$    & $24$      & $41.5692$  & $0.076250(10)$ \\            
        \textsc{fcc}-$1,3$          & $36$    & $36$      & $53.5692$  & $0.066084(6)$ \\
        \textsc{fcc}-$2,3$          & $30$    & $30$      & $50.0545$  & $0.068875(5)$ \\
        \textsc{fcc}-$1,2,3$        & $42$    & $42$      & $62.0545$  & $0.061884(6)$\tnote{d} \\
        \textsc{fcc}-$4$            & $12$    & $12$      & $24$  & $0.19924(6)$\tnote{e} \\      
        \textsc{fcc}-$1,4$          & $24$    & $24$      & $36$  & $0.085553(7)$ \\
        \textsc{fcc}-$2,4$          & $18$    & $18$      & $32.4853$  & $0.13730(10)$\tnote{f} \\
        \textsc{fcc}-$3,4$          & $36$    & $36$      & $65.5692$  & $0.055337(7)$ \\
        \textsc{fcc}-$1,2,4$        & $30$    & $30$      & $44.4853$  & $0.075148(8)$ \\
        \textsc{fcc}-$1,3,4$        & $48$    & $48$      & $77.5692$ & $0.051326(6)$ \\
        \textsc{fcc}-$2,3,4$        & $42$    & $42$      & $74.0545$ & $0.051743(7)$ \\
        \textsc{fcc}-$1,2,3,4$      & $54$    & $54$     & $86.0545$ & $0.048778(8)$\tnote{g} \\
        \textsc{fcc}-$5$            & $24$    & $24$      & $53.6656$  & $0.063758(8)$ \\
        \textsc{fcc}-$1,5$          & $36$    & $36$      & $65.6656$  & $0.053561(9)$ \\
        \textsc{fcc}-$2,5$          & $30$    & $30$      & $62.1509$  & $0.057435(5)$ \\
        \textsc{fcc}-$3,5$          & $48$    & $48$      & $95.2349$ & $0.041826(6)$ \\
        \textsc{fcc}-$4,5$          & $36$    & $36$      & $77.6656$ & $0.048517(7)$ \\
        \textsc{fcc}-$1,2,5$        & $42$    & $42$      & $74.1509$ & $0.050508(8)$ \\
        \textsc{fcc}-$1,3,5$        & $60$    & $60$     & $107.2349$ & $0.039540(10)$ \\
        \textsc{fcc}-$1,4,5$        & $48$    & $48$      & $89.6656$ & $0.043701(9)$ \\
        \textsc{fcc}-$2,3,5$        & $54$    & $54$     & $103.7201$ & $0.039806(6)$ \\
        \textsc{fcc}-$2,4,5$        & $42$    & $42$      & $86.1509$ & $0.045201(9)$ \\
        \textsc{fcc}-$3,4,5$        & $60$    & $60$     & $119.2349$ & $0.036542(8)$ \\
        \textsc{fcc}-$1,2,3,5$      & $66$    & $66$     & $115.7201$ & $0.038224(6)$ \\
        \textsc{fcc}-$1,2,4,5$      & $54$    & $54$     & $98.1509$ & $0.041724(6)$ \\
        \textsc{fcc}-$1,3,4,5$      & $72$    & $72$     & $131.2349$ & $0.035229(9)$ \\
        \textsc{fcc}-$2,3,4,5$      & $66$    & $66$     & $127.7201$ & $0.035080(10)$ \\
        \textsc{fcc}-$1,2,3,4,5$    & $78$    & $78$     & $139.7201$ & $0.034156(6)$ \\
	\hline\hline
	\end{tabular}
    \begin{tablenotes}
    \footnotesize
    \item[a] $0.1992365(10)$ \cite{LorenzZiff98}, $0.19923517(20)$ \cite{XuWangLvDeng14}.
    \item[b] $0.31160768(15)$ \cite{XuWangLvDeng14}, $0.3116060(48)$ \cite{KozaPola16}, $0.3116077(2)$ \cite{WangZhouZhangGaroniDeng13}, $0.3116081(13)$ \cite{BallesterosFernandezMartin-MayorParisiRuiz-Lorenzo99}.
    \item[c] $0.15950(12)$ \cite{Malarz15}.
    \item[d] $0.0618842(8)$ \cite{XuWangHuDeng21}.
    \item[e] $0.1992365(10)$ \cite{LorenzZiff98}, $0.19923517(20)$ \cite{XuWangLvDeng14}.
    \item[f] $0.1373045(5)$ \cite{XunHaoZiff21}, $0.1372(1)$ \cite{KurzawskiMalarz12}.
    \item[g] $0.0500(5)$ \cite{GawronCieplak91}.
    \end{tablenotes}
	\label{tab04}
    \end{threeparttable}
\end{table}	

\begin{table}[htbp]
\centering
\caption{Site percolation thresholds for objects of various complex neighborhoods on body centered cubic lattice. The object name encodes the involved coordination zones ($i\le5$), to which sites in the neighborhood belong. The coordination number $z$, the weighted distance $\xi$, and the weighted coordination number $\zeta$ are also presented. Here $r$ was normalized to ensure that $r_1 = 1$.}
    \begin{threeparttable}
	\begin{tabular}{ccccc}
	\hline\hline
		neighborhood                & $z$       &$\xi$      & $\zeta$      & $p_c$(site) \\
		\hline
		\textsc{bcc}-$1$            & $8$       & $8$       & $8$          & $0.24596(6)$\tnote{a} \\       
		\textsc{bcc}-$2$            & $6$       & $4$       & $6.9282$     & $0.31161(9)$\tnote{b} \\
        \textsc{bcc}-$1,2$          & $14$      & $12$      & $14.9282$    & $0.17594(6)$\tnote{c} \\   
        \textsc{bcc}-$3$            & $12$      & $10.6667$ & $19.5959$    & $0.19924(6)$\tnote{d} \\              
        \textsc{bcc}-$1,3$          & $20$      & $18.6667$ & $27.5959$    & $0.11048(8)$ \\     
        \textsc{bcc}-$2,3$          & $18$      & $14.6667$ & $26.5241$    & $0.13730(10)$\tnote{e} \\
        \textsc{bcc}-$1,2,3$        & $26$      & $22.6667$ & $34.5241$    & $0.095908(8)$\tnote{f} \\
        \textsc{bcc}-$4$            & $24$      & $22$      & $45.9565$    & $0.071243(7)$ \\
        \textsc{bcc}-$1,4$          & $32$      & $30$      & $53.9565$    & $0.062735(5)$ \\
        \textsc{bcc}-$2,4$          & $30$      & $26$      & $52.8847$    & $0.065777(7)$ \\
        \textsc{bcc}-$3,4$          & $36$      & $32.6667$ & $65.5524$    & $0.057510(10)$ \\
        \textsc{bcc}-$1,2,4$        & $38$      & $34$      & $60.8847$    & $0.060218(8)$ \\
        \textsc{bcc}-$1,3,4$        & $44$      & $40.6667$ & $73.5524$    & $0.054243(7)$ \\
        \textsc{bcc}-$2,3,4$        & $42$      & $36.6667$ & $72.4806$    & $0.054379(9)$ \\
        \textsc{bcc}-$1,2,3,4$      & $50$      & $44.6667$ & $80.4806$    & $0.052164(6)$ \\
        \textsc{bcc}-$5$            & $8$       & $6.4$     & $16$         & $0.24596(6)$\tnote{g} \\     
        \textsc{bcc}-$1,5$          & $16$      & $14.4$    & $24$         & $0.11255(5)$ \\
        \textsc{bcc}-$2,5$          & $14$      & $10.4$    & $22.9282$    & $0.13615(5)$\tnote{h} \\
        \textsc{bcc}-$3,5$          & $20$      & $17.0667$ & $35.5959$    & $0.10376(6)$\tnote{i} \\
        \textsc{bcc}-$4,5$          & $32$      & $28.4$    & $61.9565$    & $0.059076(6)$ \\
        \textsc{bcc}-$1,2,5$        & $22$      & $18.4$    & $30.9282$    & $0.095133(7)$ \\
        \textsc{bcc}-$1,3,5$        & $28$      & $25.0667$ & $43.5959$    & $0.075452(8)$ \\  
        \textsc{bcc}-$1,4,5$        & $40$      & $36.4$    & $69.9565$    & $0.054340(10)$ \\
        \textsc{bcc}-$2,3,5$        & $26$      & $21.0667$ & $42.5241$    & $0.097644(6)$\tnote{j} \\
        \textsc{bcc}-$2,4,5$        & $38$      & $32.4$    & $68.8847$    & $0.055115(5)$ \\
        \textsc{bcc}-$3,4,5$        & $44$      & $39.0667$ & $81.5524$    & $0.049312(8)$ \\
        \textsc{bcc}-$1,2,3,5$      & $34$      & $29.0667$ & $50.5241$    & $0.071642(8)$ \\
        \textsc{bcc}-$1,2,4,5$      & $46$      & $40.4$    & $76.8847$    & $0.052071(9)$ \\
        \textsc{bcc}-$1,3,4,5$      & $52$      & $47.0667$ & $89.5524$    & $0.047440(10)$ \\
        \textsc{bcc}-$2,3,4,5$      & $50$      & $43.0667$ & $88.4806$    & $0.047575(5)$ \\
        \textsc{bcc}-$1,2,3,4,5$    & $58$      & $51.0667$ & $96.4806$    & $0.046177(7)$ \\
	\hline\hline
	\end{tabular}
    \begin{tablenotes}
    \footnotesize
    \item[a] $0.2455(1)$ \cite{KurzawskiMalarz12}.
    \item[b] $0.31160768(15)$ \cite{XuWangLvDeng14}, $0.3116060(48)$ \cite{KozaPola16}, $0.3116077(2)$ \cite{WangZhouZhangGaroniDeng13}, $0.3116081(13)$ \cite{BallesterosFernandezMartin-MayorParisiRuiz-Lorenzo99}.
    \item[c] $0.20490(12)$ \cite{Malarz15}.
    \item[d] $0.1992365(10)$ \cite{LorenzZiff98}, $0.19923517(20)$ \cite{XuWangLvDeng14}.
    \item[e] $0.1373045(5)$ \cite{XunHaoZiff21}, $0.1372(1)$ \cite{KurzawskiMalarz12}.
    \item[f] $0.0959084(6)$ \cite{XunHaoZiff21}.
    \item[g] $0.2455(1)$ \cite{KurzawskiMalarz12}.
    \item[h] $0.1420(1)$ \cite{KurzawskiMalarz12}.
    \item[i] $0.1036(1)$ \cite{KurzawskiMalarz12}.
    \item[j] $0.0976445(10)$, $0.0976444(6)$ \cite{XunHaoZiff21}, $0.0976(1)$ \cite{KurzawskiMalarz12}.
    \end{tablenotes}
	\label{tab05}
    \end{threeparttable}
\end{table}	

\begin{figure}[htbp]
\centering
\includegraphics[width=3.3in]{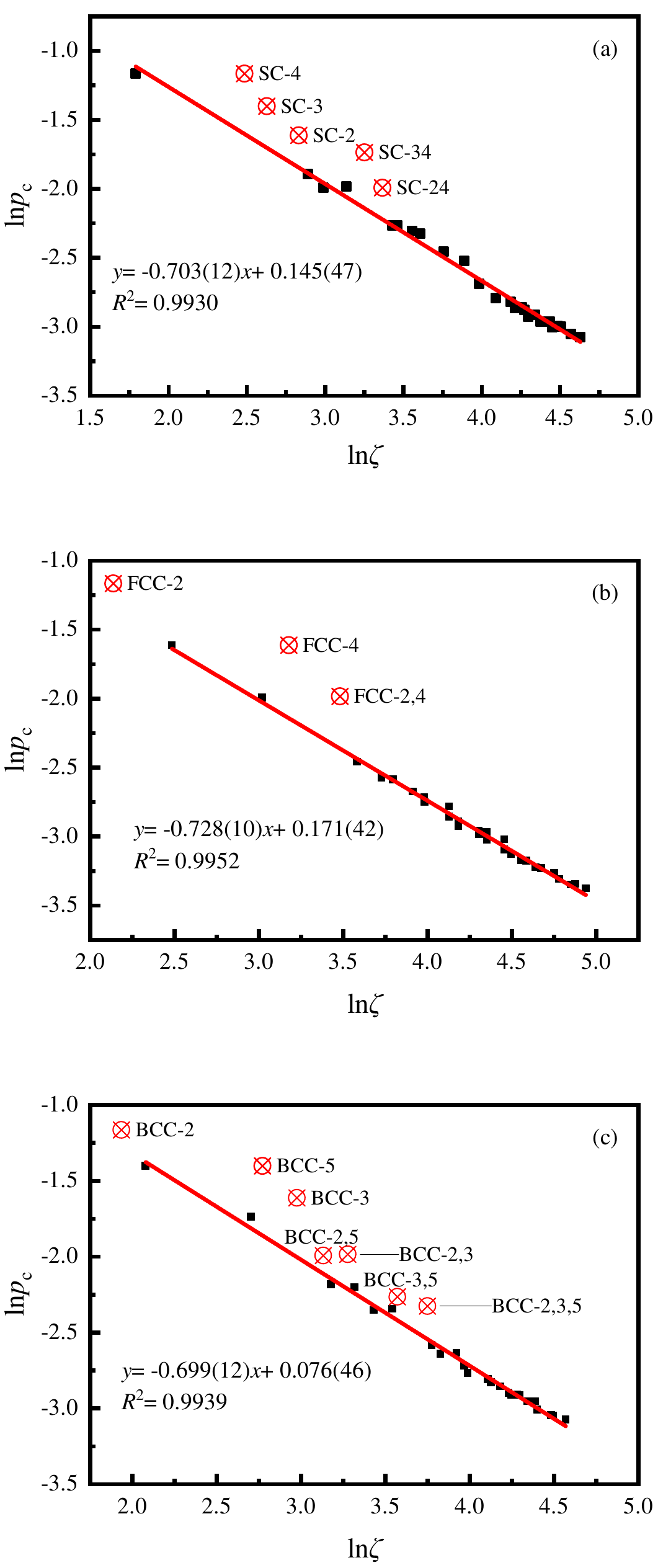}
\caption{Plots of $\ln p_c$ versus $\ln \zeta$ for the site percolation thresholds of complex neighborhoods extending to the fifth coordination zone on the simple cubic lattice (a), face centered cubic lattice (b), and body centered cubic lattice (c), respectively. Inflated neighborhoods marked with $\protect\crosscircle$ (also collected in the footnotes of Table III) are excluded from the fitting procedure.}
\label{SC_FCC_BCC_zetavspc}
\end{figure}

For a more detailed analysis, the plots of $\ln p_c$ versus $\ln \zeta$ are indicated in Fig.\ \ref{SC_FCC_BCC_zetavspc} (a), (b) and (c) for complex neighborhoods on SC, FCC and BCC lattices, respectively. The approximately linear dependence of $\ln p_c$ on $\ln \zeta$ indicates the power-law behavior $p_c \propto \zeta^{-\gamma}$ described by Eq.\ (\ref{eq:zeta}). An inflated neighborhood is obtained by uniformly rescaling all neighbor displacement vectors of a smaller neighborhood. Such a rescaling partitions the lattice into equivalent sublattices, each having the same connectivity as the original system, and therefore leaves $p_c$ unchanged. However, $\zeta$ changes because it depends explicitly on the neighbor distances. For example, \textsc{sc}-$4$ is an inflated form of \textsc{sc}-$1$ with twice the lattice constant, so the two neighborhoods have the same $p_c$ but different values of $\zeta$. Including both the original and inflated neighborhoods would introduce multiple scale-related representations of the same connectivity structure into the fit. We therefore exclude the inflated neighborhoods listed in footnotes aa–ii of Table \ref{tab03} from the fitting. The same treatment method was adopted by Malarz \cite{Malarz24}. Then the fitting results for each lattice are displayed as follows.
\\

\ \ \ \ SC: $y = -0.703(12)x + 0.145(47)$.
\\

\ \ \ \ FCC: $y = -0.728(10)x + 0.171(42)$.
\\

\ \ \ \ BCC: $y = -0.699(12)x + 0.076(46)$.
\\

\noindent{These fittings imply that the exponents $\gamma_{\mathrm{SC}}^{\zeta} = 0.703(12)$, $\gamma_{\mathrm{FCC}}^{\zeta} = 0.728(10)$ and $\gamma_{\mathrm{BCC}}^{\zeta} = 0.699(12)$ for SC, FCC and BCC lattice, respectively.} 

\begin{figure}[htbp]
\centering
\includegraphics[width=3.4in]{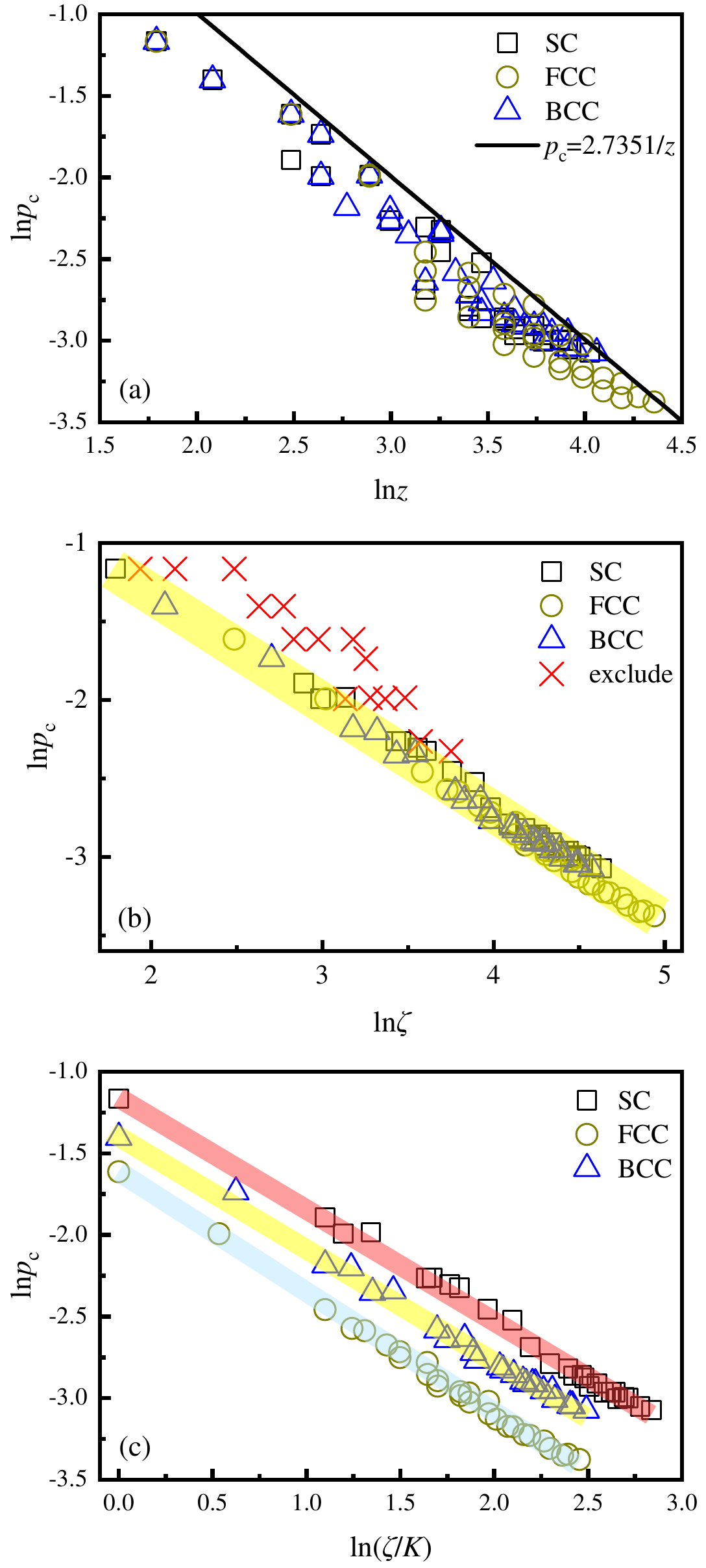}
\caption{Site percolation thresholds $p_c$ for complex neighborhoods containing sites up to the fifth coordination zone on simple cubic ($\square$), face centered cubic ($\bigcirc$), and body centered cubic ($\triangle$) lattices as dependent on (a) coordination number $z$, (b) index $\zeta$, and (c) index $\zeta/K$. The solid line in (a) corresponds to $p_c = (2^d \eta_c)/z$ with the continuum threshold for spheres $\eta_c = 0.34189$ \cite{LorenzZiff01,TorquatoJiao12,GoriTrombettoni15}, and points marked with crosses ($\times$) in (b) correspond to inflated neighborhoods which are excluded from the fitting procedure.}
\label{3comb_z_zeta_zetaK_vs_pc}
\end{figure}

In Fig.\ \ref{3comb_z_zeta_zetaK_vs_pc}, we present the site percolation thresholds $p_c$ for neighborhoods containing sites up to the fifth coordination zone versus (a) total coordination number $z$, (b) index $\zeta$, and (c) index $\zeta / K$, on SC ($\square$), FCC ($\bigcirc$), and BCC ($\triangle$) lattices. The number $K$ indicates lattice connectivity, that is, $K = 6$, $12$, and $8$ for the SC, FCC, and BCC lattices, respectively. As mentioned in Tables \ref{tab03}, \ref{tab04}, \ref{tab05}, strong $p_c(z)$ degeneration is observed, as shown in Fig.\ \ref{3comb_z_zeta_zetaK_vs_pc}(a). By contrast, the introduction of index $\zeta$ can effectively eliminate this degeneracy, leading to a nearly perfect separation of $p_c$ [see Fig.\ \ref{3comb_z_zeta_zetaK_vs_pc}(b)]. In Fig.\ \ref{3comb_z_zeta_zetaK_vs_pc}(b), the inflated neighborhoods discussed above are marked by crosses ($\times$), and excluded from the fit. Then it is observed that the data points for various lattice models collapse nicely onto a single curve, indicating a kind of universality. The linear fit of the data gives
\\

\ \ \ \ $y = -0.719(9)x + 0.1637(35)$.
\\

\noindent{This fitting implies the power law $p_c \propto \zeta^{-\gamma}$ with exponent $\gamma_{\mathrm{3D}}^{\zeta} = 0.719(9)$.} As shown in Fig.\ \ref{3comb_z_zeta_zetaK_vs_pc}(c), dividing the index $\zeta$ by $K$ (for viewing) separates the lattice-specific curves that collapse in panel (b), and fits to the data for each lattice yield the values reported in Fig.\ \ref{SC_FCC_BCC_zetavspc}.

In addition, the solid line in Fig.\ \ref{3comb_z_zeta_zetaK_vs_pc}(a) represents the relation of Eq.\ (\ref{eq:zpcsite}), $p_c = (2^d \eta_c)/z \sim 2.7351/z$, with the continuum threshold for spheres $\eta_c = 0.34189$ \cite{LorenzZiff01,TorquatoJiao12,GoriTrombettoni15}. It can be seen that all the results we have obtained do not exceed this line. This implies the existence of an upper critical limit for the percolation threshold, that is $p_c \le (2^d \eta_c)/z$. This finding generalizes the two-dimensional results reported in Ref.\ \cite{CiepluchaUtnickiWoloszynMalarz25}. Regrettably, we have not been able to identify the corresponding lower critical limit.

\section{Conclusions}
\label{conclusions}

In this work, we performed high-precision Monte Carlo simulations of percolation on the SC, FCC, and BCC lattices with extended-range neighborhoods. We systematically investigated two classes of neighborhoods, namely, cubic neighborhoods and complex (non-compact) neighborhoods. Using an efficient single-cluster growth algorithm, we obtained high-precision estimates of the percolation threshold $p_c$ for approximately 100 distinct structures. For cubic neighborhoods, we considered sizes ranging from $2 \times 2 \times 2$ to $22 \times 22 \times 22$, corresponding to coordination numbers $26\le z\le12166$. For bond percolation, the product $zp_c$ approaches the Bethe-lattice limit $z p_c \to 1$ as $z$ increases, with the finite-$z$ corrections following $zp_c-1\sim z^{-x}$, where $x\approx0.65$, consistent with $x = (d-1)/d$ given by Xu et al \cite{XuWangHuDeng21}. More precisely, the behavior for large $z$ is found to follow \cite{OuyangDengBlote18}
\begin{equation}
    zp_{c}\sim 1 + a_1 z^{-1/2} + a_2 z^{-1} + a_3 z^{-3/2}.
\label{eq:zpc}
\end{equation}
This expression describes the finite-$z$ corrections to the mean-field (Bethe-lattice) behavior. In two dimensions, data fitting of our former simulation results \cite{XunHaoZiff2026} leads to $x = 0.53-0.54$. The slight difference between our measured exponent and $x = 1/2$ is presumably due to higher-order corrections. In three dimensions, our numerical estimate $x = 0.65$ is closer to the theoretical value $x = 2/3$. This comparison demonstrates that as the system dimensionality increases, the higher-order corrections will gradually diminish.

For site percolation of cubic neighborhoods, a qualitatively different convergence behavior was found. The value of $zp_c$ first increases and then decreases as $z$ increases. It is not until the value of $z$ becomes sufficiently large that a good linear variation pattern emerges. For $z \ge 3374$, linear fitting leads to $zp_c = 2.5987(1)$, predicting a value of continuum percolation threshold of aligned cubes to be $\eta_c = 0.324838(12)$ from the perspective of $zp_c = 2^{d} \eta_c$.

In non-compact neighborhoods, extensive simulations on site percolation were carried out with neighborhoods extending to the fifth coordination zone on SC, FCC and BCC lattices. Due to their porous structure, the relationship between the thresholds $p_c$ and $z$ exhibits degeneracy. To resolve this degeneracy, three indices of the coordination number $z$, the weighted distance $\xi = \sum_i z_i r_i^2 /i$ , and the weighted coordination number $\zeta = \sum_i z_i r_i$ are attempted, and results of the comparison indicate that $\zeta$ is effective in removing the degeneracy. Besides, we found that all measured thresholds collapse onto a single power law $p_c \propto \zeta^{-\gamma}$ with $\gamma_\mathrm{3D}^{\zeta} = 0.719(9)$, independent of the underlying lattice. When the index $\zeta$ is divided by $K$, which denotes the lattice connectivity, the collapsed data separate into lattice-specific curves. Fits to these curves yield the power-law dependence $p_c \propto (\zeta / K)^{-\gamma}$ with $\gamma_{\mathrm{SC}}^{\zeta} = 0.703(12)$, $\gamma_{\mathrm{FCC}}^{\zeta} = 0.728(10)$, and $\gamma_{\mathrm{BCC}}^{\zeta} = 0.699(12)$ for the SC, FCC, and BCC lattices, respectively. 


\section{Acknowledgments}
The authors would like to thank Professor Robert M. Ziff for his guidance on this work. This work was funded by the National Natural Science Foundation of China (Nos. 12275002, 12347169), the Natural Science Foundation of Jiangsu Province (No. BK20240892).

\section{Data availability}
The data that support the findings of this article are included in the Supplemental Material as plots \cite{ZhangFanYangXun26supplementary} and data and codes can be found in our Github repository \cite{github}. The actual
raw data can be obtained from the authors upon reasonable
request.

\bibliographystyle{unsrt}
\bibliography{bibliography.bib}

\end{document}


\title{Supplementary Material: Extended-range percolation on lattices with cubic and non-compact neighborhoods in three dimensions}

\author{Zhe Zhang}
\affiliation{School of Physics Science and Technology, Yangzhou University, Yangzhou, 225009, China}

\author{Yuxuan Fan}
\affiliation{School of Physics Science and Technology, Yangzhou University, Yangzhou, 225009, China}

\author{Yang Yang}
\affiliation{School of Material Sciences and Physics, China University of Mining and Technology, Xuzhou 221116, China}

\author{Yusong Tu}
\affiliation{School of Physics Science and Technology, Yangzhou University, Yangzhou, 225009, China}

\author{Zhipeng Xun}
\email{zpxun@cumt.edu.cn}
\affiliation{School of Material Sciences and Physics, China University of Mining and Technology, Xuzhou 221116, China}

\date{\today}

\begin{abstract} 
Supplementary material for Zhe Zhang, Yuxuan Fan, Yang Yang, Yusong Tu, and Zhipeng Xun, Phys.\ Rev.\ E (2026).
\end{abstract}

\maketitle

\begin{figure*}[htbp]
\centering
\includegraphics[width=7.2in]{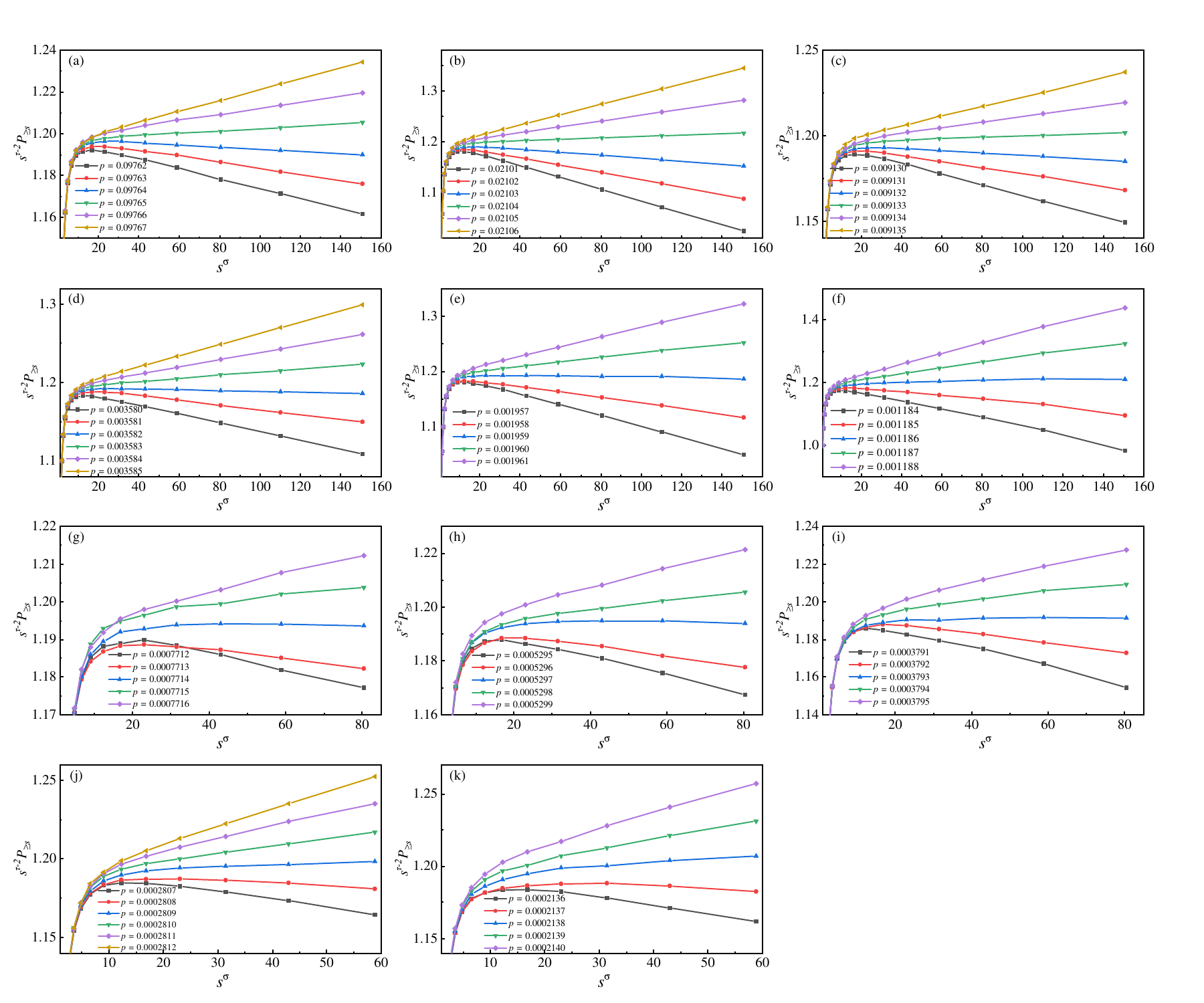}
\caption{Plot of $s^{\tau-2}P_{\geq{s}}$ versus $s^{\sigma}$ with $\tau=2.18905$ and $\sigma=0.4522$ for site percolation of the \textsc{cub}-(2) (a), \textsc{cub}-(4) (b), \textsc{cub}-(6) (c), \textsc{cub}-(8) (d), \textsc{cub}-(10) (e), \textsc{cub}-(12) (f), \textsc{cub}-(14) (g), \textsc{cub}-(16) (h), \textsc{cub}-(18) (i), \textsc{cub}-(20) (j) and \textsc{cub}-(22) (k) objects under different values of $p$. For each objects, as $p \rightarrow p_c$, the value of $s^{\tau-2}P_{\geq{s}}$ becomes horizontal for large $s$.} 
\label{site_sigma_33} 
\end{figure*}

\begin{figure*}[htbp]
\centering
\includegraphics[width=7.2in]{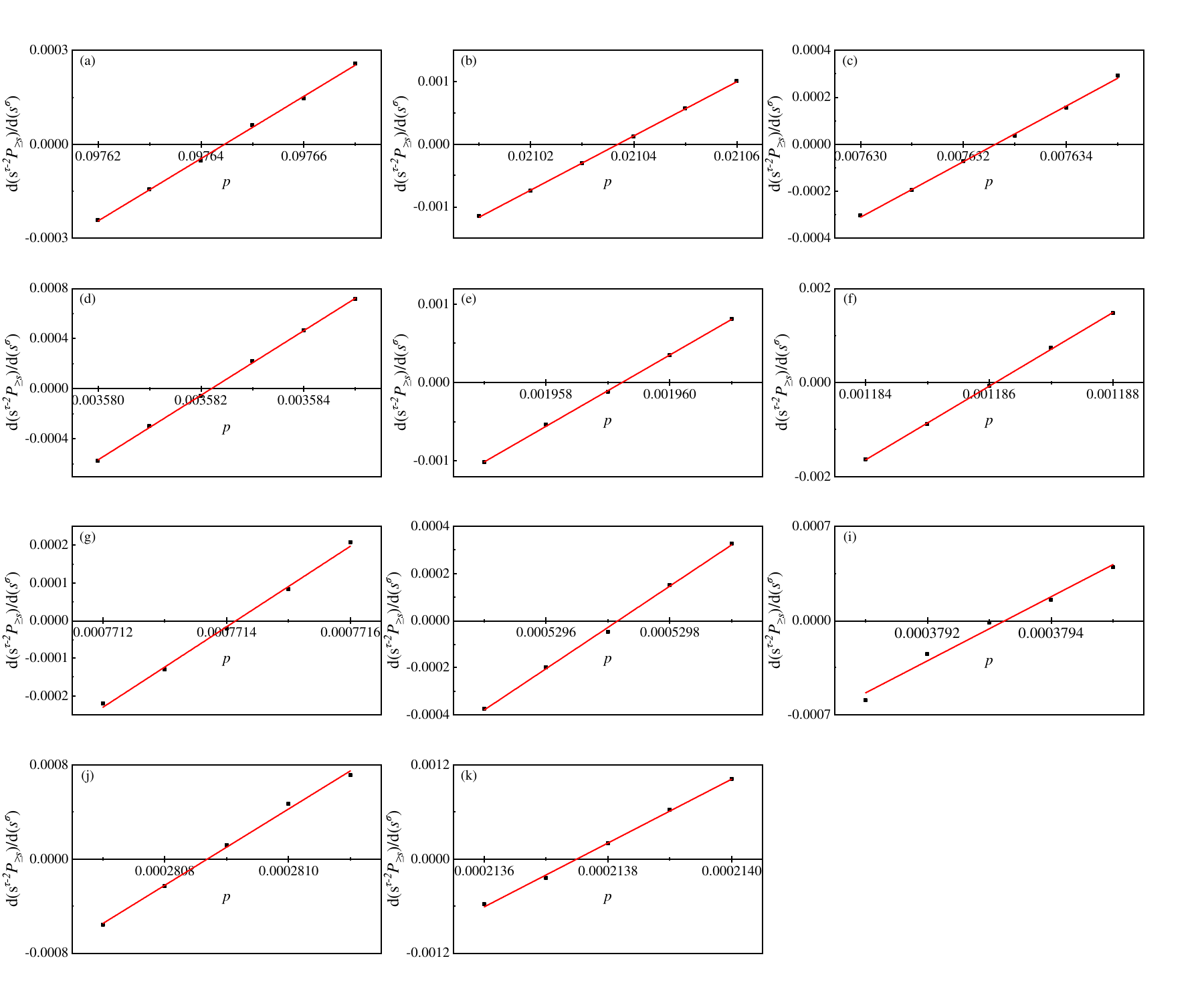}
\caption{Plot of $\mathrm{d}(s^{\tau-2}P_{\geq{s}})/\mathrm{d}(s^{\sigma})$ versus $p$ for site percolation of the \textsc{cub}-(2) (a), \textsc{cub}-(4) (b), \textsc{cub}-(6) (c), \textsc{cub}-(8) (d), \textsc{cub}-(10) (e), \textsc{cub}-(12) (f), \textsc{cub}-(14) (g), \textsc{cub}-(16) (h), \textsc{cub}-(18) (i), \textsc{cub}-(20) (j) and \textsc{cub}-(22) (k) objects. The central value of $p_c$ can be estimated from the $p$ intercept.}
\label{site_sigma_diff_33} 
\end{figure*}

\begin{figure*}[htbp] 
\centering
\includegraphics[width=7.2in]{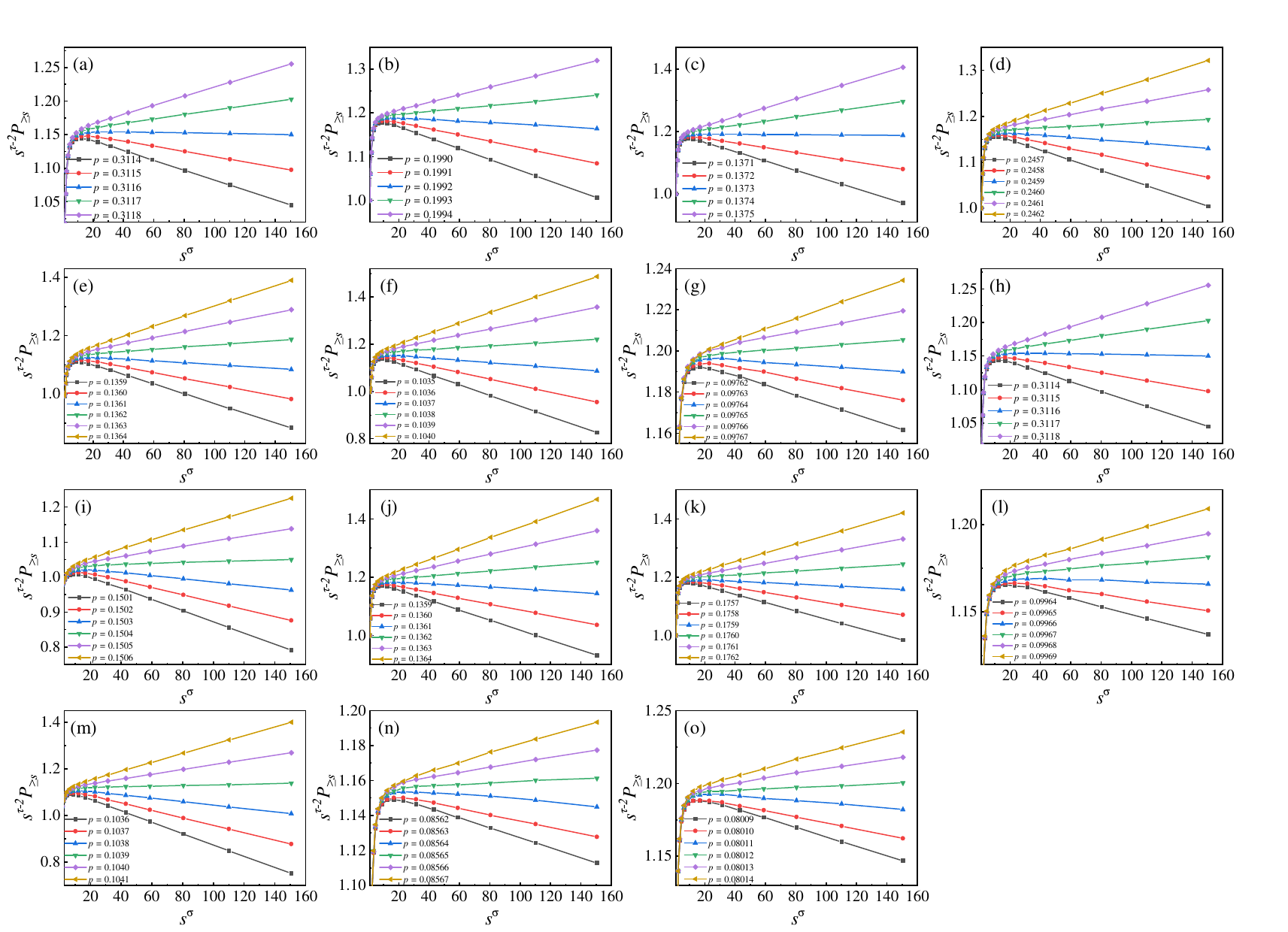}
\caption{Plot of $s^{\tau-2}P_{\geq s}$ versus $s^{\sigma}$ with $\tau = 2.18905$ and $\sigma = 0.4522$ for site percolation of the \textsc{sc}-1 (a), \textsc{sc}-2 (b), \textsc{sc}-1,2 (c), \textsc{sc}-3 (d), \textsc{sc}-1,3 (e), \textsc{sc}-2,3 (f), \textsc{sc}-1,2,3 (g), \textsc{sc}-4 (h), \textsc{sc}-1,4 (i), \textsc{sc}-2,4 (j), \textsc{sc}-3,4 (k), \textsc{sc}-1,2,4 (l), \textsc{sc}-1,3,4 (m), \textsc{sc}-2,3,4 (n) and \textsc{sc}-1,2,3,4 (o) objects on simple cubic lattice under different values of $p$. For each object, as $p \rightarrow p_c$, the value of $s^{\tau-2}P_{\geq s}$ becomes horizontal for large $s$.}
\label{sc_NN4_sigma}
\end{figure*}

\begin{figure*}[htbp] 
\centering
\includegraphics[width=7.2in]{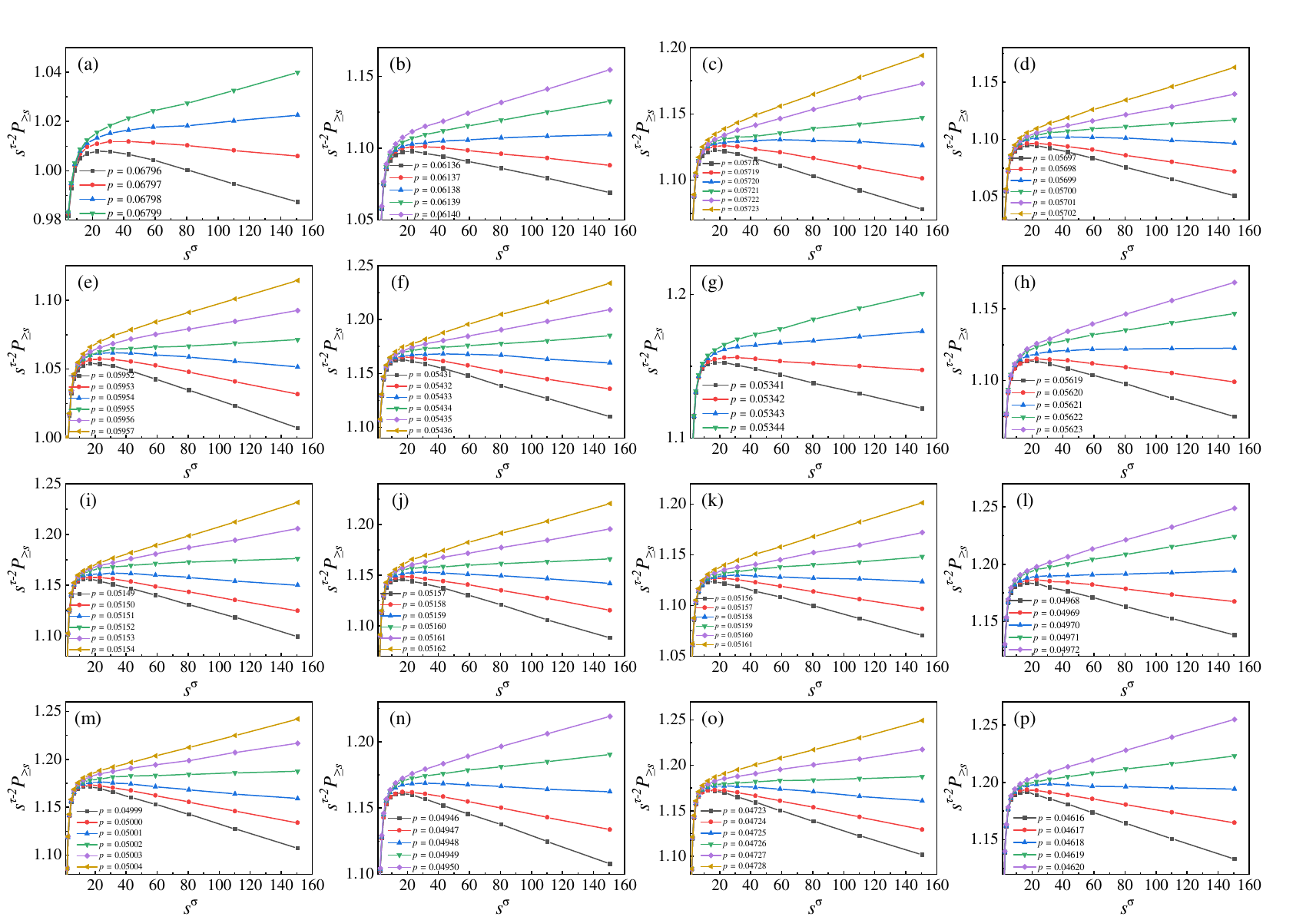}
\caption{Plot of $s^{\tau-2}P_{\geq s}$ versus $s^{\sigma}$ with $\tau = 2.18905$ and $\sigma = 0.4522$ for site percolation of the \textsc{sc}-$5$ (a), \textsc{sc}-$1,5$ (b), \textsc{sc}-$2,5$ (c), \textsc{sc}-$3,5$ (d), \textsc{sc}-$4,5$ (e), \textsc{sc}-$1,2,5$ (f), \textsc{sc}-$1,3,5$ (g), \textsc{sc}-$1,4,5$ (h), \textsc{sc}-$2,3,5$ (i), \textsc{sc}-$2,4,5$ (j), \textsc{sc}-$3,4,5$ (k), \textsc{sc}-$1,2,3,5$ (l), \textsc{sc}-$1,2,4,5$ (m), \textsc{sc}-$1,3,4,5$ (n), \textsc{sc}-$2,3,4,5$ (o) and \textsc{sc}-$1,2,3,4,5$ (p) objects on simple cubic lattice under different values of $p$. For each object, as $p \rightarrow p_c$, the value of $s^{\tau-2}P_{\geq s}$ becomes horizontal for large $s$.}
\label{sc_NN5_sigma}
\end{figure*}

\begin{figure*}[htbp] 
\centering
\includegraphics[width=7.2in]{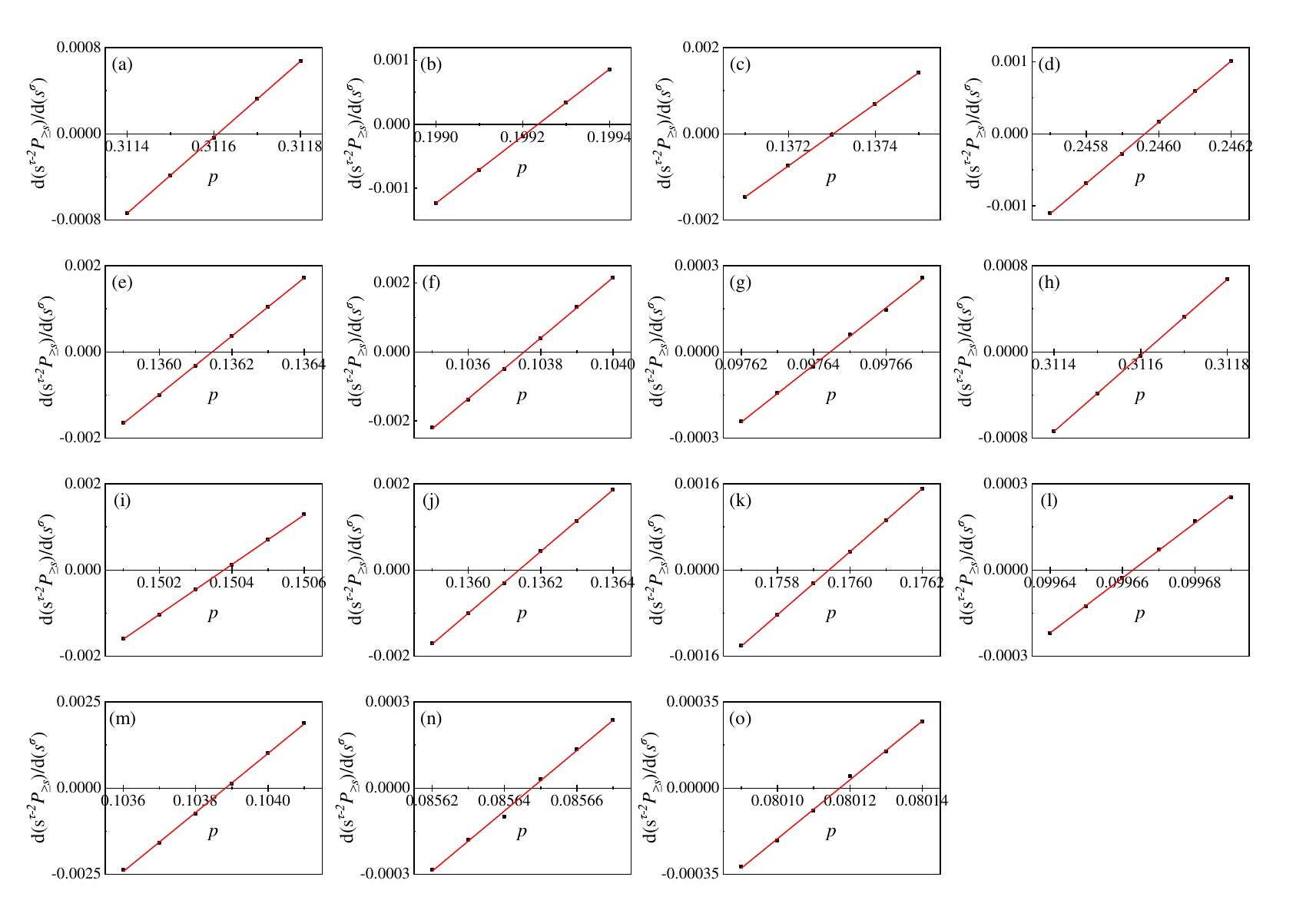}
\caption{Plot of the slope $\mathrm{d}(s^{\tau-2}P_{\geq s}) / \mathrm{d}(s^{\sigma})$ versus $p$ for site percolation of the \textsc{sc}-1 (a), \textsc{sc}-2 (b), \textsc{sc}-1,2 (c), \textsc{sc}-3 (d), \textsc{sc}-1,3 (e), \textsc{sc}-2,3 (f), \textsc{sc}-1,2,3 (g), \textsc{sc}-4 (h), \textsc{sc}-1,4 (i), \textsc{sc}-2,4 (j), \textsc{sc}-3,4 (k), \textsc{sc}-1,2,4 (l), \textsc{sc}-1,3,4 (m), \textsc{sc}-2,3,4 (n) and \textsc{sc}-1,2,3,4 (o) objects on simple cubic lattice. The central value of $p_c$ can be estimated from the $p$ intercept along the abscissa.}
\label{sc_NN4_sigma_slope}
\end{figure*}

\begin{figure*}[htbp] 
\centering
\includegraphics[width=7.2in]{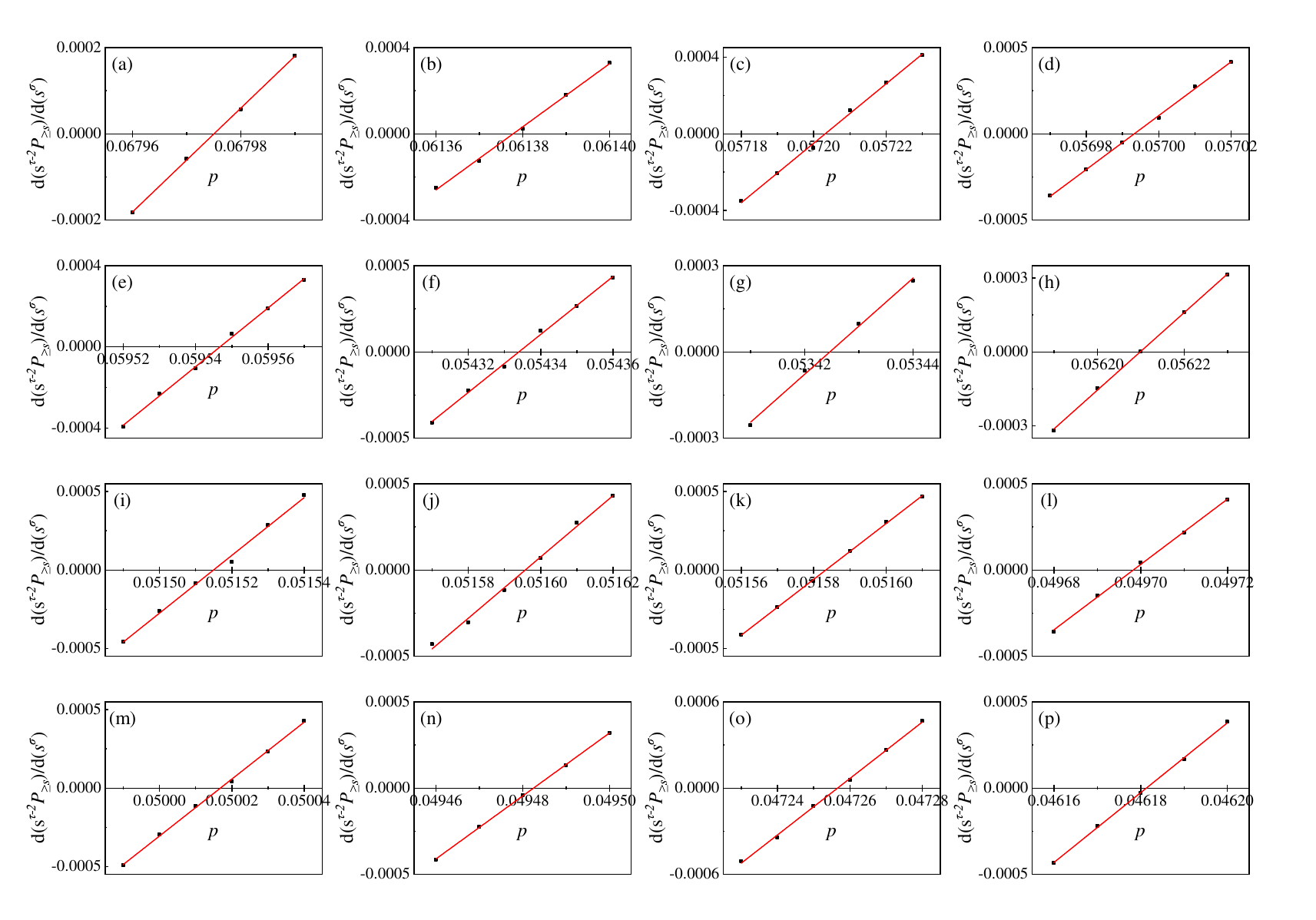}
\caption{Plot of the slope $\mathrm{d}(s^{\tau-2}P_{\geq s}) / \mathrm{d}(s^{\sigma})$ versus $p$ for site percolation of the \textsc{sc}-$5$ (a), \textsc{sc}-$1,5$ (b), \textsc{sc}-$2,5$ (c), \textsc{sc}-$3,5$ (d), \textsc{sc}-$4,5$ (e), \textsc{sc}-$1,2,5$ (f), \textsc{sc}-$1,3,5$ (g), \textsc{sc}-$1,4,5$ (h), \textsc{sc}-$2,3,5$ (i), \textsc{sc}-$2,4,5$ (j), \textsc{sc}-$3,4,5$ (k), \textsc{sc}-$1,2,3,5$ (l), \textsc{sc}-$1,2,4,5$ (m), \textsc{sc}-$1,3,4,5$ (n), \textsc{sc}-$2,3,4,5$ (o) and \textsc{sc}-$1,2,3,4,5$ (p) objects on simple cubic lattice. The central value of $p_c$ can be estimated from the $p$ intercept along the abscissa.}
\label{sc_NN5_sigma_slope}
\end{figure*}

\begin{figure*}[htbp] 
\centering
\includegraphics[width=7.2in]{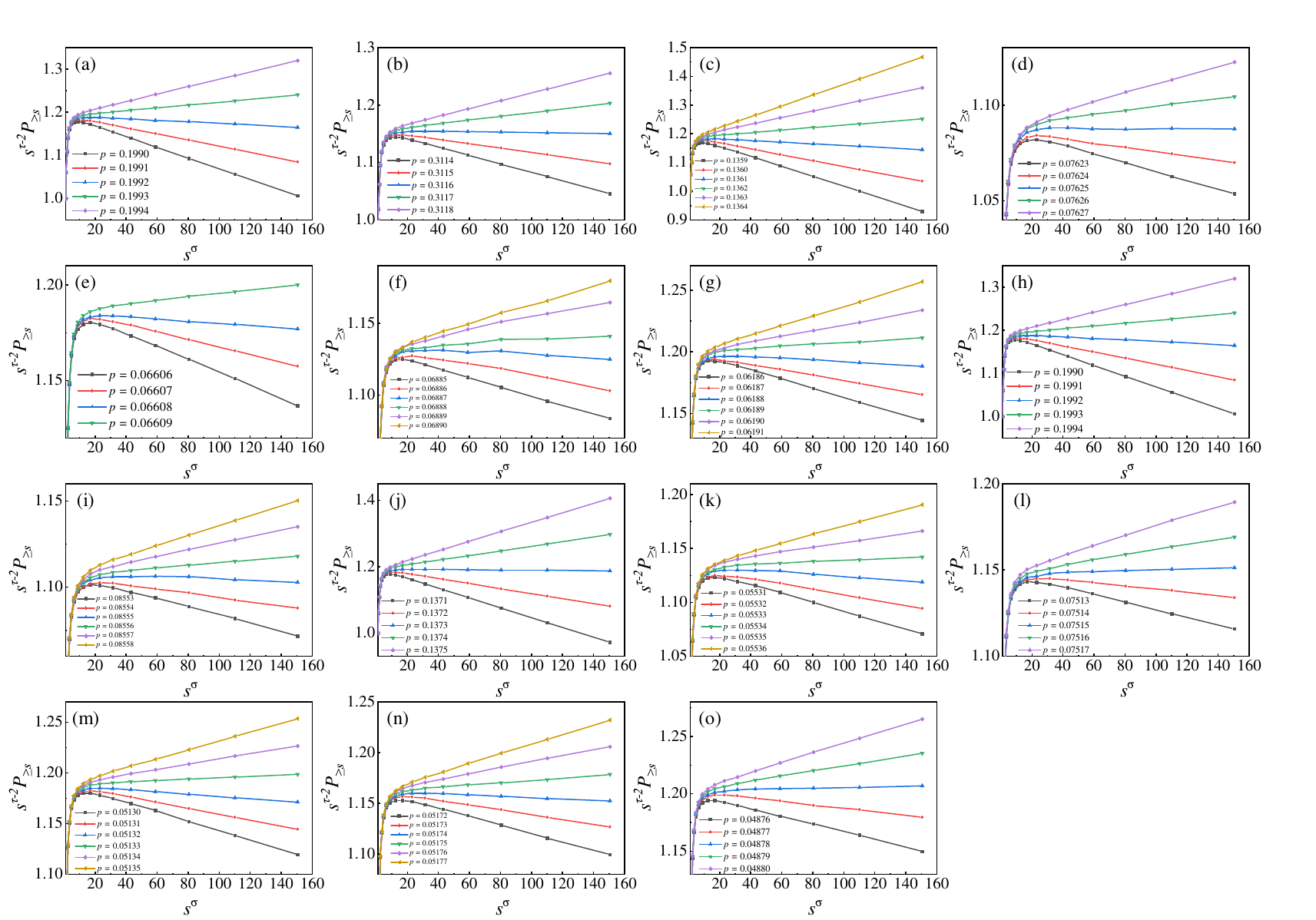}
\caption{Plot of $s^{\tau-2}P_{\geq s}$ versus $s^{\sigma}$ with $\tau = 2.18905$ and $\sigma = 0.4522$ for site percolation of the \textsc{fcc}-1 (a), \textsc{fcc}-2 (b), \textsc{fcc}-1,2 (c), \textsc{fcc}-3 (d), \textsc{fcc}-1,3 (e), \textsc{fcc}-2,3 (f), \textsc{fcc}-1,2,3 (g), \textsc{fcc}-4 (h), \textsc{fcc}-1,4 (i), \textsc{fcc}-2,4 (j), \textsc{fcc}-3,4 (k), \textsc{fcc}-1,2,4 (l), \textsc{fcc}-1,3,4 (m), \textsc{fcc}-2,3,4 (n) and \textsc{fcc}-1,2,3,4 (o) objects on face centered cubic lattice under different values of $p$. For each object, as $p \rightarrow p_c$, the value of $s^{\tau-2}P_{\geq s}$ becomes horizontal for large $s$.}
\label{fcc_NN4_sigma}
\end{figure*}

\begin{figure*}[htbp] 
\centering
\includegraphics[width=7.2in]{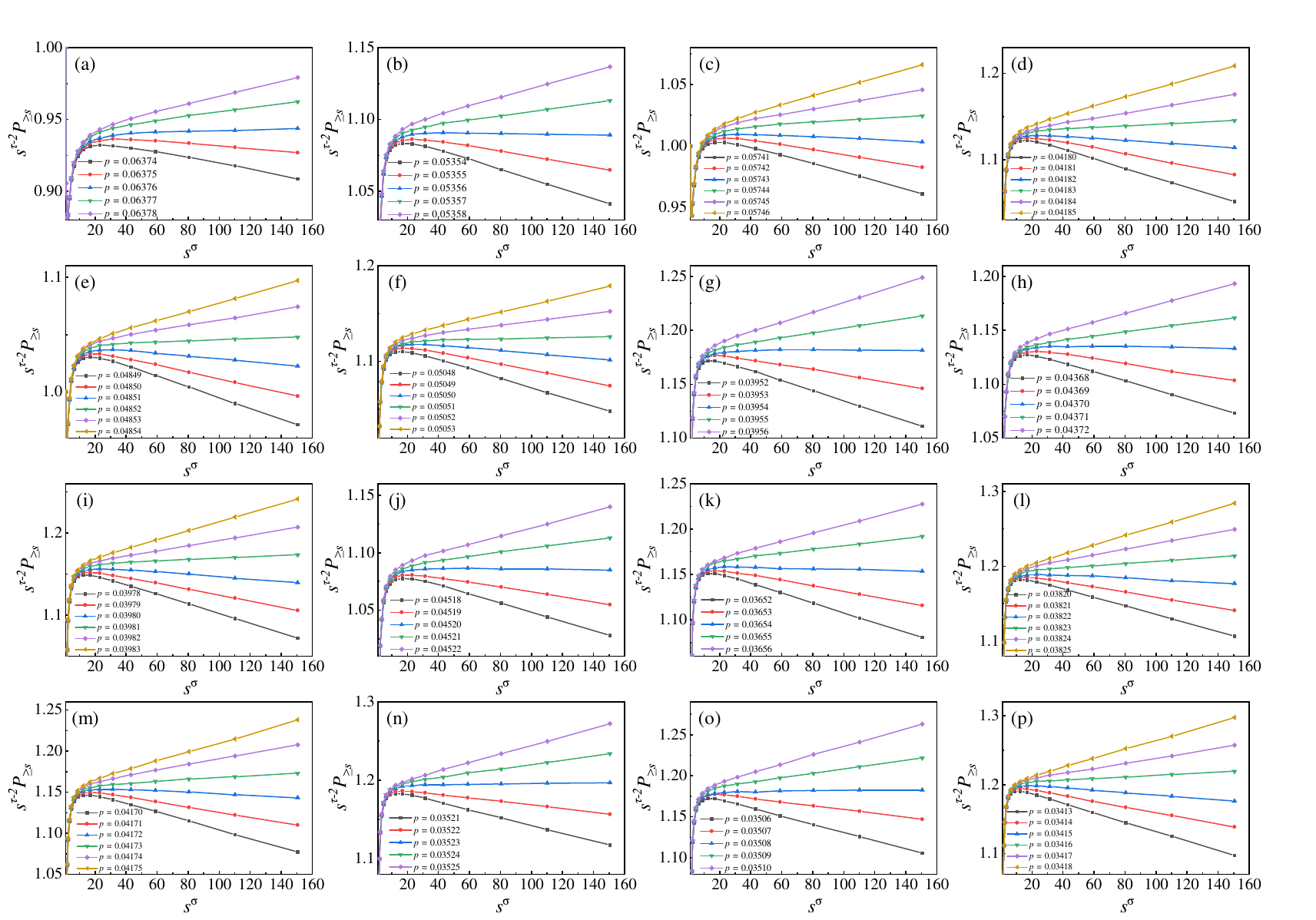}
\caption{Plot of $s^{\tau-2}P_{\geq s}$ versus $s^{\sigma}$ with $\tau = 2.18905$ and $\sigma = 0.4522$ for site percolation of the \textsc{fcc}-$5$ (a), \textsc{fcc}-$1,5$ (b), \textsc{fcc}-$2,5$ (c), \textsc{fcc}-$3,5$ (d), \textsc{fcc}-$4,5$ (e), \textsc{fcc}-$1,2,5$ (f), \textsc{fcc}-$1,3,5$ (g), \textsc{fcc}-$1,4,5$ (h), \textsc{fcc}-$2,3,5$ (i), \textsc{fcc}-$2,4,5$ (j), \textsc{fcc}-$3,4,5$ (k), \textsc{fcc}-$1,2,3,5$ (l), \textsc{fcc}-$1,2,4,5$ (m), \textsc{fcc}-$1,3,4,5$ (n), \textsc{fcc}-$2,3,4,5$ (o) and \textsc{fcc}-$1,2,3,4,5$ (p) objects on face centered cubic lattice under different values of $p$. For each object, as $p \rightarrow p_c$, the value of $s^{\tau-2}P_{\geq s}$ becomes horizontal for large $s$.}
\label{fcc_NN5_sigma}
\end{figure*}

\begin{figure*}[htbp] 
\centering
\includegraphics[width=7.2in]{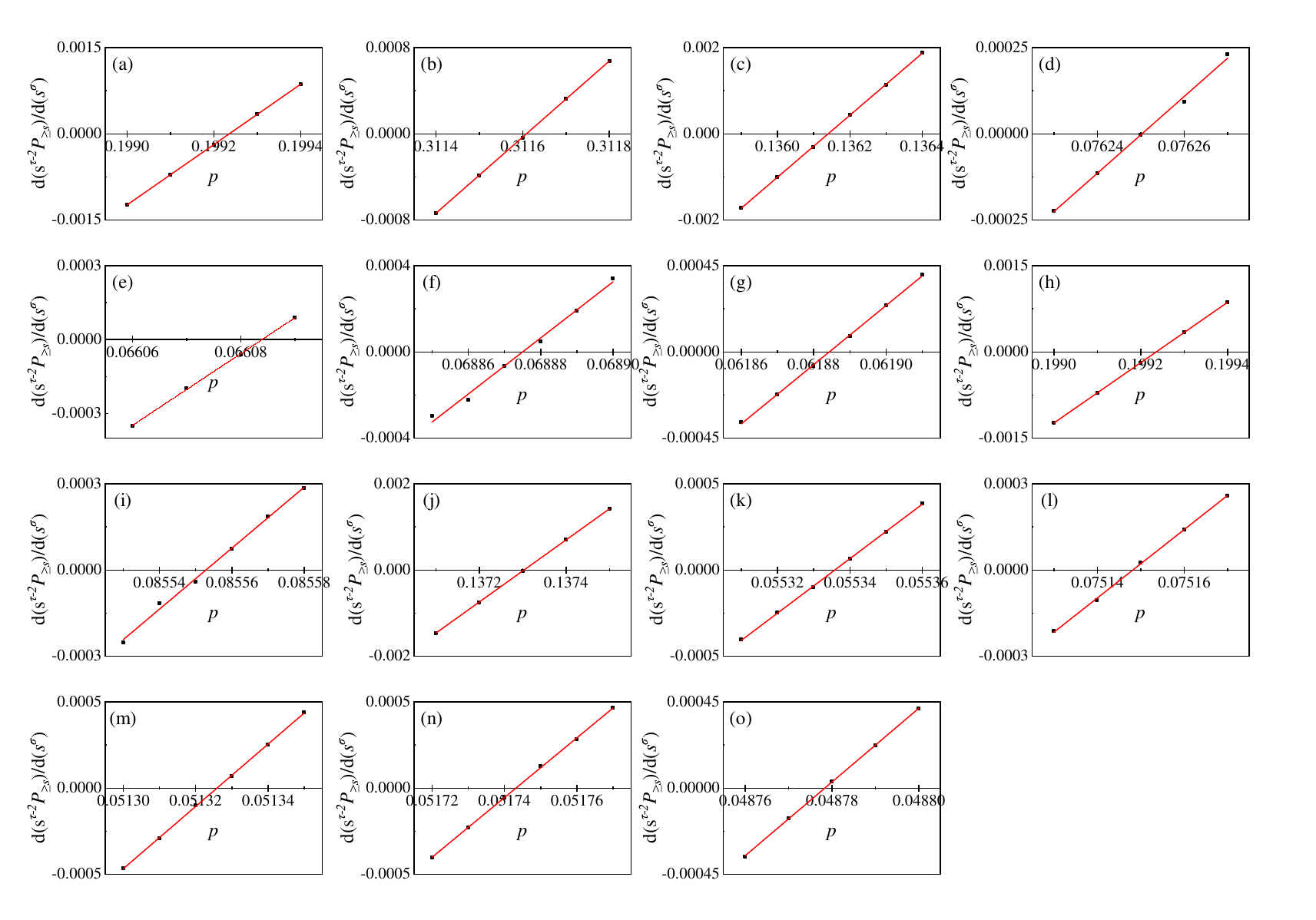}
\caption{Plot of the slope $\mathrm{d}(s^{\tau-2}P_{\geq s}) / \mathrm{d}(s^{\sigma})$ versus $p$ for site percolation of the \textsc{fcc}-1 (a), \textsc{fcc}-2 (b), \textsc{fcc}-1,2 (c), \textsc{fcc}-3 (d), \textsc{fcc}-1,3 (e), \textsc{fcc}-2,3 (f), \textsc{fcc}-1,2,3 (g), \textsc{fcc}-4 (h), \textsc{fcc}-1,4 (i), \textsc{fcc}-2,4 (j), \textsc{fcc}-3,4 (k), \textsc{fcc}-1,2,4 (l), \textsc{fcc}-1,3,4 (m), \textsc{fcc}-2,3,4 (n) and \textsc{fcc}-1,2,3,4 (o) objects on face centered cubic lattice. The central value of $p_c$ can be estimated from the $p$ intercept along the abscissa.}
\label{fcc_NN4_sigma_slope}
\end{figure*}

\begin{figure*}[htbp] 
\centering
\includegraphics[width=7.2in]{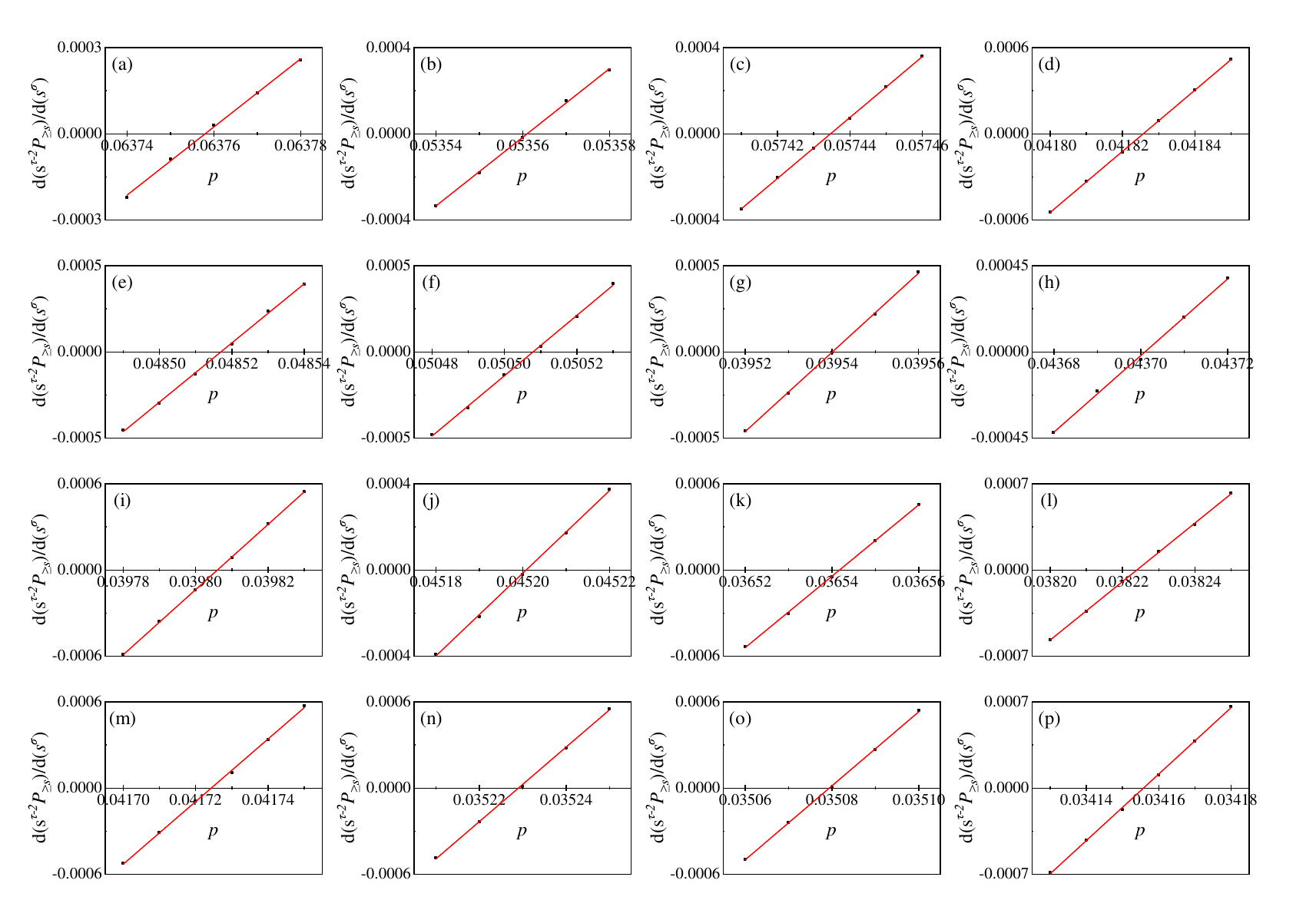}
\caption{Plot of the slope $\mathrm{d}(s^{\tau-2}P_{\geq s}) / \mathrm{d}(s^{\sigma})$ versus $p$ for site percolation of the \textsc{fcc}-$5$ (a), \textsc{fcc}-$1,5$ (b), \textsc{fcc}-$2,5$ (c), \textsc{fcc}-$3,5$ (d), \textsc{fcc}-$4,5$ (e), \textsc{fcc}-$1,2,5$ (f), \textsc{fcc}-$1,3,5$ (g), \textsc{fcc}-$1,4,5$ (h), \textsc{fcc}-$2,3,5$ (i), \textsc{fcc}-$2,4,5$ (j), \textsc{fcc}-$3,4,5$ (k), \textsc{fcc}-$1,2,3,5$ (l), \textsc{fcc}-$1,2,4,5$ (m), \textsc{fcc}-$1,3,4,5$ (n), \textsc{fcc}-$2,3,4,5$ (o) and \textsc{fcc}-$1,2,3,4,5$ (p) objects on face centered cubic lattice. The central value of $p_c$ can be estimated from the $p$ intercept along the abscissa.}
\label{fcc_NN5_sigma_slope}
\end{figure*}

\begin{figure*}[htbp] 
\centering
\includegraphics[width=7.2in]{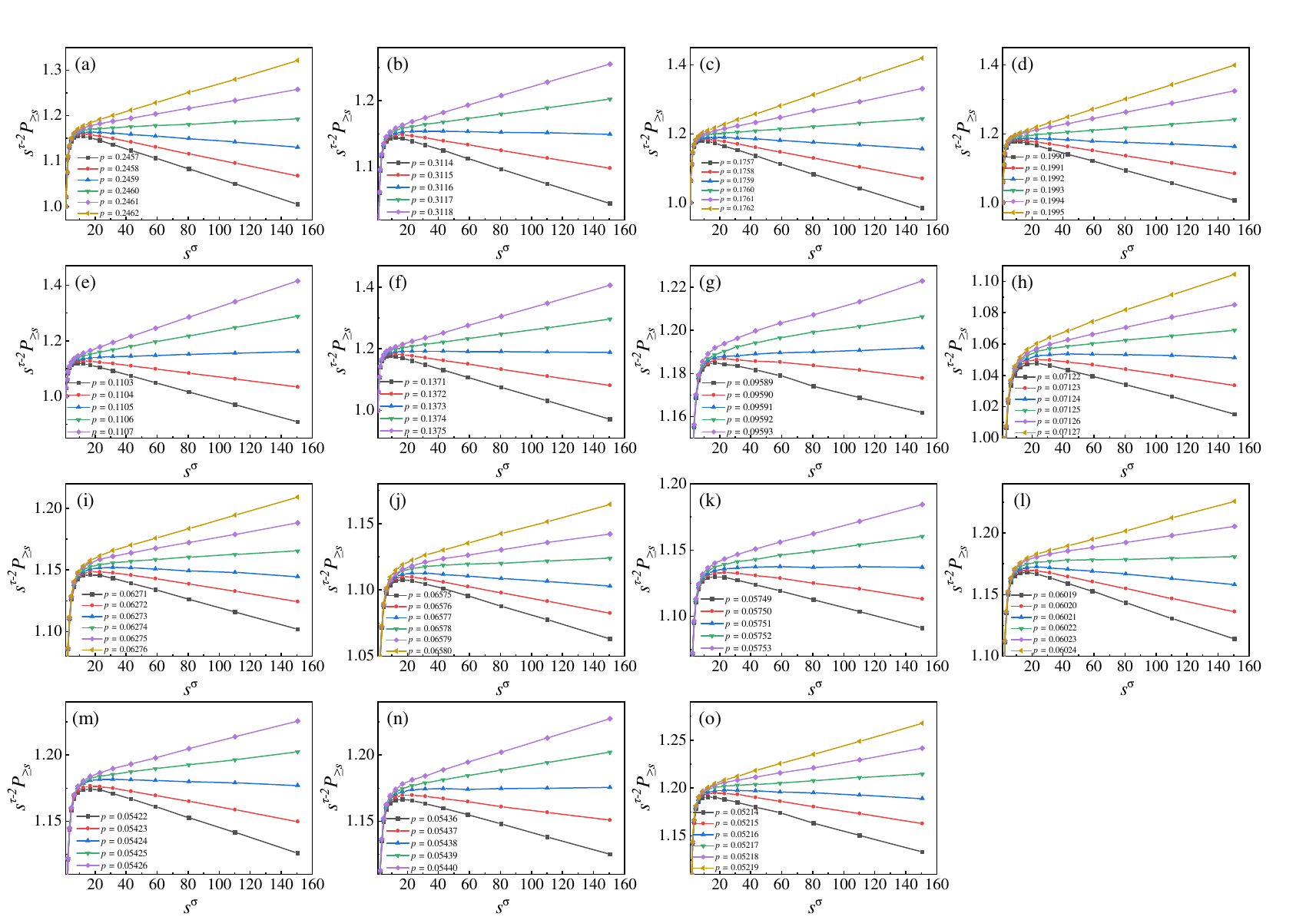}
\caption{Plot of $s^{\tau-2}P_{\geq s}$ versus $s^{\sigma}$ with $\tau = 2.18905$ and $\sigma = 0.4522$ for site percolation of the \textsc{bcc}-1 (a), \textsc{bcc}-2 (b), \textsc{bcc}-1,2 (c), \textsc{bcc}-3 (d), \textsc{bcc}-1,3 (e), \textsc{bcc}-2,3 (f), \textsc{bcc}-1,2,3 (g), \textsc{bcc}-4 (h), \textsc{bcc}-1,4 (i), \textsc{bcc}-2,4 (j), \textsc{bcc}-3,4 (k), \textsc{bcc}-1,2,4 (l), \textsc{bcc}-1,3,4 (m), \textsc{bcc}-2,3,4 (n) and \textsc{bcc}-1,2,3,4 (o) objects on body centered cubic lattice under different values of $p$. For each object, as $p \rightarrow p_c$, the value of $s^{\tau-2}P_{\geq s}$ becomes horizontal for large $s$.}
\label{bcc_NN4_sigma}
\end{figure*}

\begin{figure*}[htbp] 
\centering
\includegraphics[width=7.2in]{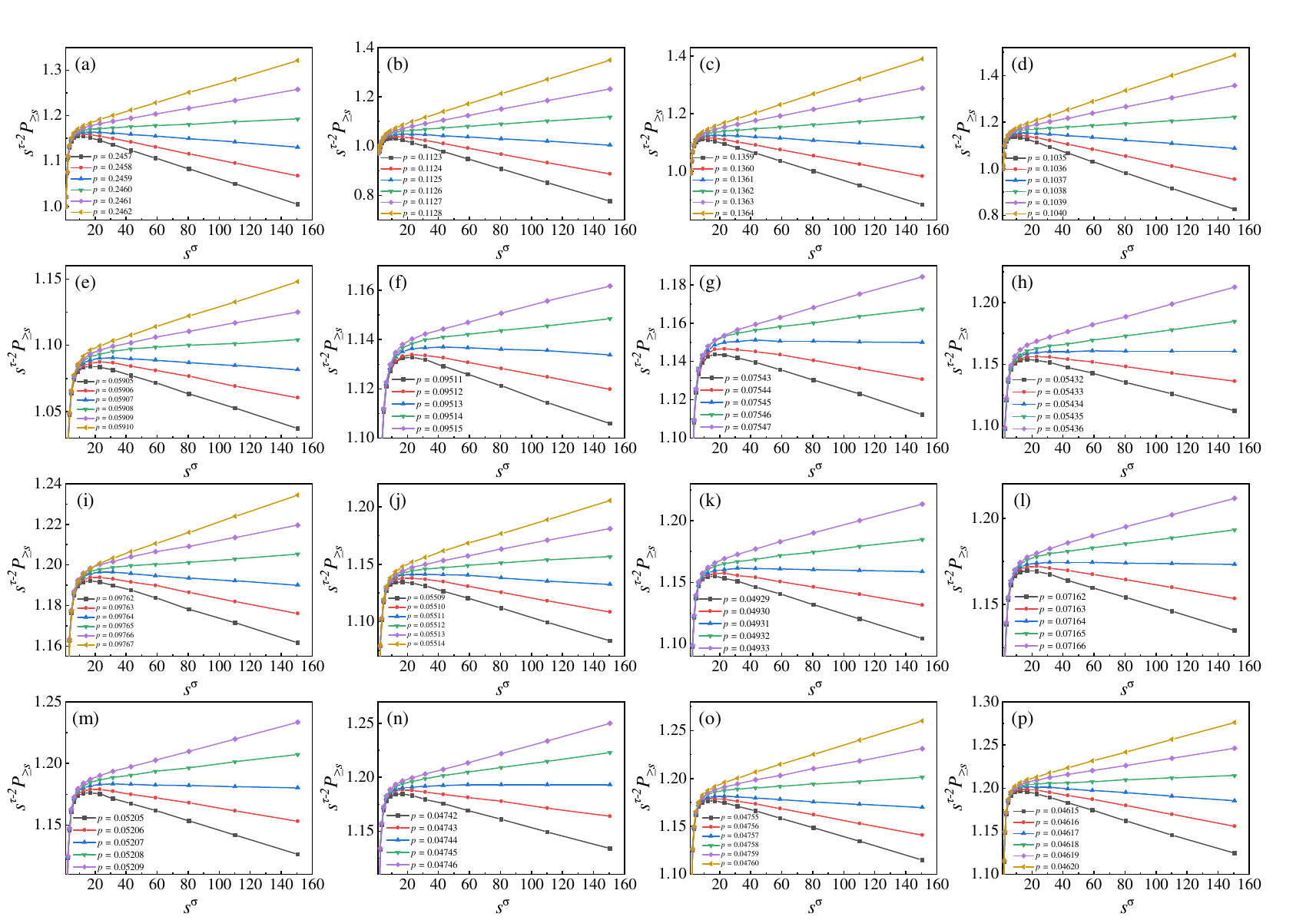}
\caption{Plot of $s^{\tau-2}P_{\geq s}$ versus $s^{\sigma}$ with $\tau = 2.18905$ and $\sigma = 0.4522$ for site percolation of the \textsc{bcc}-$5$ (a), \textsc{bcc}-$1,5$ (b), \textsc{bcc}-$2,5$ (c), \textsc{bcc}-$3,5$ (d), \textsc{bcc}-$4,5$ (e), \textsc{bcc}-$1,2,5$ (f), \textsc{bcc}-$1,3,5$ (g), \textsc{bcc}-$1,4,5$ (h), \textsc{bcc}-$2,3,5$ (i), \textsc{bcc}-$2,4,5$ (j), \textsc{bcc}-$3,4,5$ (k), \textsc{bcc}-$1,2,3,5$ (l), \textsc{bcc}-$1,2,4,5$ (m), \textsc{bcc}-$1,3,4,5$ (n), \textsc{bcc}-$2,3,4,5$ (o) and \textsc{bcc}-$1,2,3,4,5$ (p) objects on body centered cubic lattice under different values of $p$. For each object, as $p \rightarrow p_c$, the value of $s^{\tau-2}P_{\geq s}$ becomes horizontal for large $s$.}
\label{bcc_NN5_sigma}
\end{figure*}

\begin{figure*}[htbp] 
\centering
\includegraphics[width=7.2in]{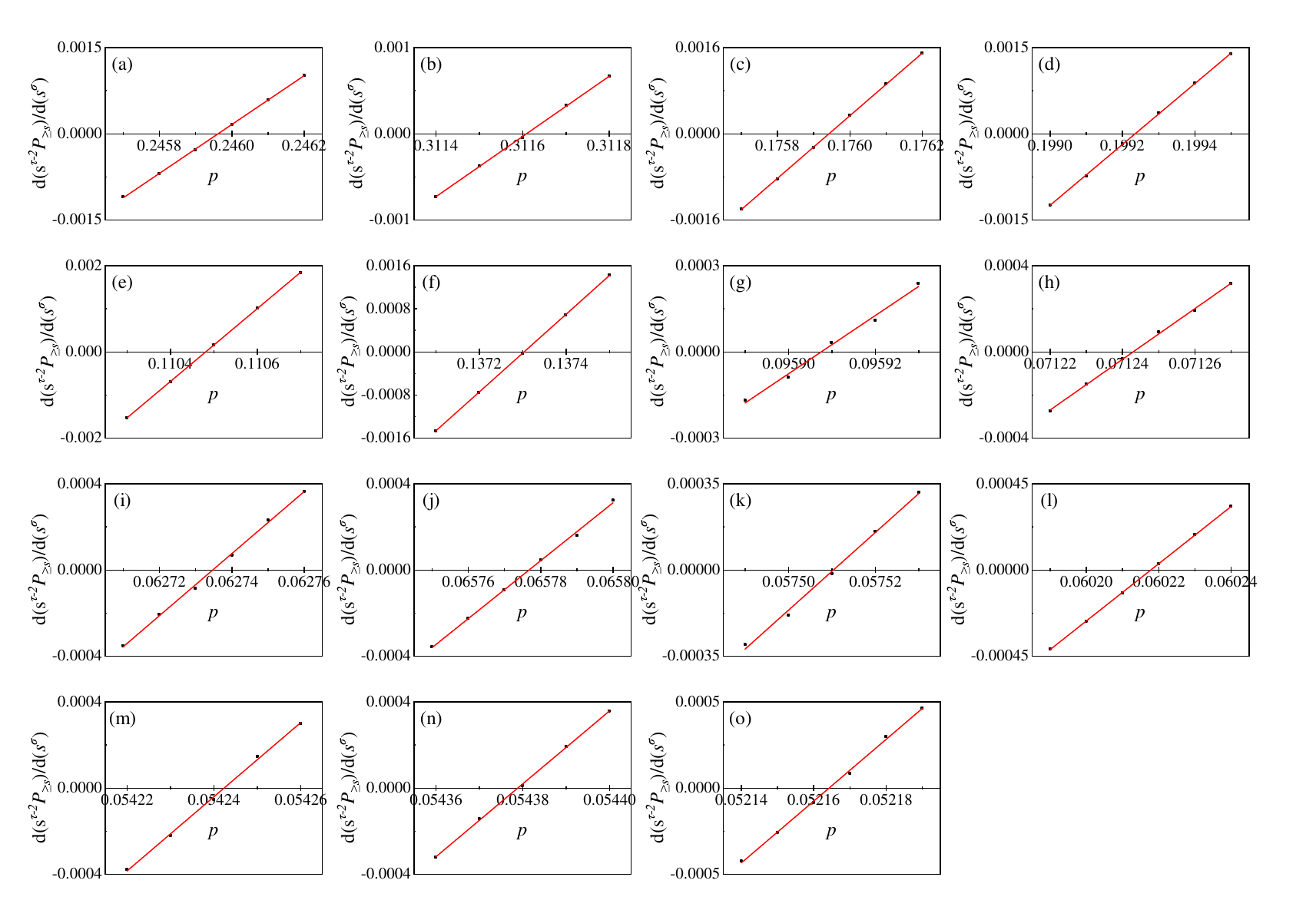}
\caption{Plot of the slope $\mathrm{d}(s^{\tau-2}P_{\geq s}) / \mathrm{d}(s^{\sigma})$ versus $p$ for site percolation of the \textsc{bcc}-1 (a), \textsc{bcc}-2 (b), \textsc{bcc}-1,2 (c), \textsc{bcc}-3 (d), \textsc{bcc}-1,3 (e), \textsc{bcc}-2,3 (f), \textsc{bcc}-1,2,3 (g), \textsc{bcc}-4 (h), \textsc{bcc}-1,4 (i), \textsc{bcc}-2,4 (j), \textsc{bcc}-3,4 (k), \textsc{bcc}-1,2,4 (l), \textsc{bcc}-1,3,4 (m), \textsc{bcc}-2,3,4 (n) and \textsc{bcc}-1,2,3,4 (o) objects on body centered cubic lattice. The central value of $p_c$ can be estimated from the $p$ intercept along the abscissa.}
\label{bcc_NN4_sigma_slope}
\end{figure*}

\begin{figure*}[htbp] 
\centering
\includegraphics[width=7.2in]{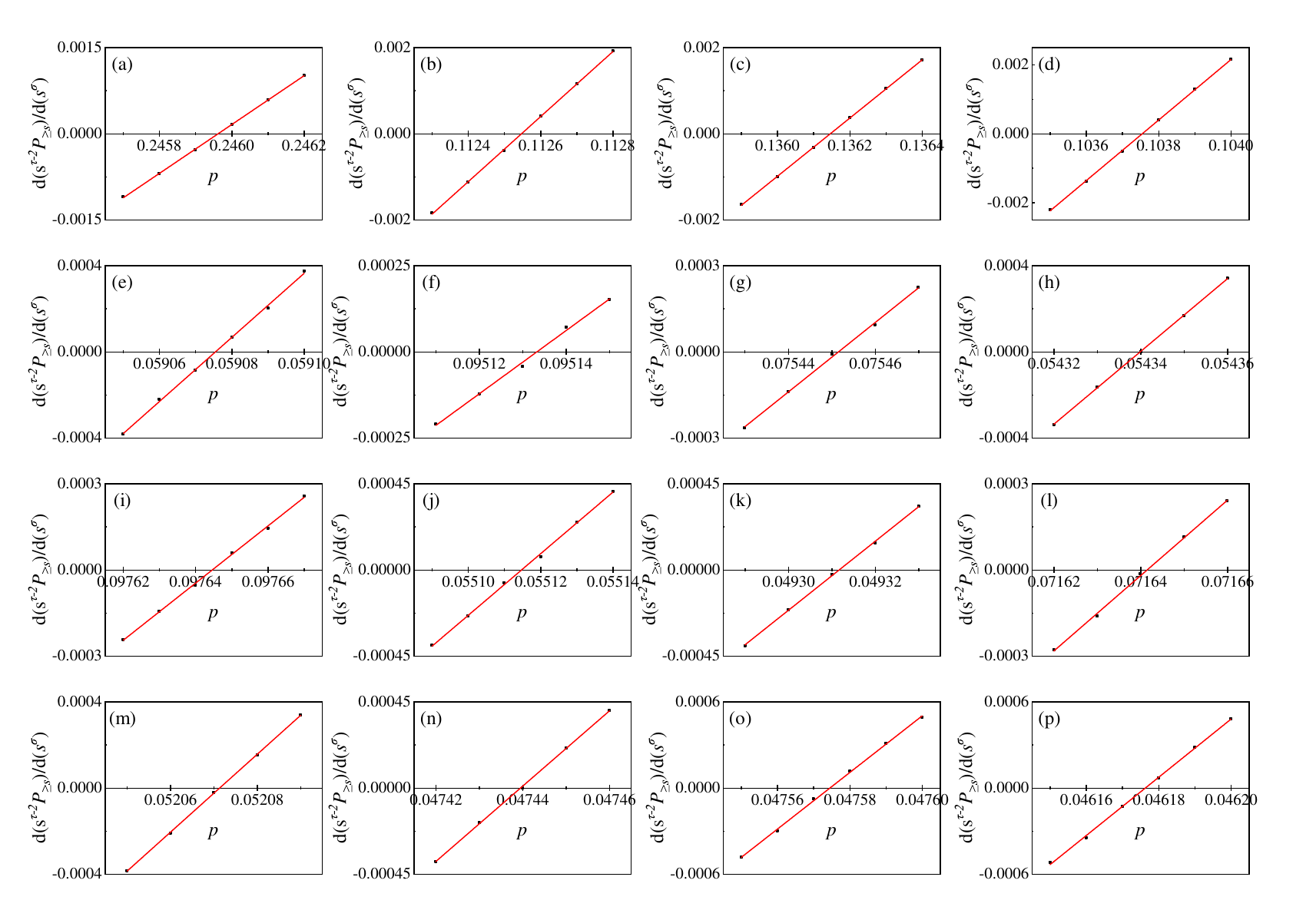}
\caption{Plot of the slope $\mathrm{d}(s^{\tau-2}P_{\geq s}) / \mathrm{d}(s^{\sigma})$ versus $p$ for site percolation of the \textsc{bcc}-$5$ (a), \textsc{bcc}-$1,5$ (b), \textsc{bcc}-$2,5$ (c), \textsc{bcc}-$3,5$ (d), \textsc{bcc}-$4,5$ (e), \textsc{bcc}-$1,2,5$ (f), \textsc{bcc}-$1,3,5$ (g), \textsc{bcc}-$1,4,5$ (h), \textsc{bcc}-$2,3,5$ (i), \textsc{bcc}-$2,4,5$ (j), \textsc{bcc}-$3,4,5$ (k), \textsc{bcc}-$1,2,3,5$ (l), \textsc{bcc}-$1,2,4,5$ (m), \textsc{bcc}-$1,3,4,5$ (n), \textsc{bcc}-$2,3,4,5$ (o) and \textsc{bcc}-$1,2,3,4,5$ (p) objects on body centered cubic lattice. The central value of $p_c$ can be estimated from the $p$ intercept along the abscissa.}
\label{bcc_NN5_sigma_slope}
\end{figure*}

\begin{itemize}
\item Cubic neighborhoods, site, Figs.~\ref{site_sigma_33} and \ref{site_sigma_diff_33},
\item Complex, SC, site, Figs.~\ref{sc_NN4_sigma}, \ref{sc_NN5_sigma}, \ref{sc_NN4_sigma_slope}, and \ref{sc_NN5_sigma_slope},
\item Complex, FCC, site, Figs.~\ref{fcc_NN4_sigma}, \ref{fcc_NN5_sigma}, \ref{fcc_NN4_sigma_slope}, and \ref{fcc_NN5_sigma_slope},
\item Complex, BCC, site, Figs.~\ref{bcc_NN4_sigma}, \ref{bcc_NN5_sigma}, \ref{bcc_NN4_sigma_slope}, and \ref{bcc_NN5_sigma_slope}.

\end{itemize}

Following are the plots of $s^{\tau-2}P_{\geq s}$ versus $s^{\sigma}$ and $\mathrm{d}(s^{\tau-2}P_{\geq s}) / \mathrm{d}(s^{\sigma})$ versus $p$ for the simulation results that are not provided in the main text.

\bibliographystyle{unsrt}
\bibliography{bibliography.bib}